\documentclass[journal]{IEEEtran}
\IEEEoverridecommandlockouts
\usepackage{cite}
\usepackage{amsmath,amssymb,amsfonts}
\usepackage{algorithmic}
\usepackage{textcomp}
\usepackage{tabularx}
\usepackage{placeins}
\usepackage[table,xcdraw]{xcolor}
\usepackage{multirow}
\usepackage{amsthm}
\usepackage{url}
\usepackage{hyperref}
\usepackage{comment}
\usepackage{tikz}
\graphicspath{ {./figure/} }
\usepackage{booktabs}
\usepackage[table,xcdraw]{xcolor}
\usepackage{algorithm, flushend}
\usepackage{subcaption}
\usepackage{marvosym}

\DeclareMathSymbol{\mathbbE}{\mathord}{AMSb}{"45}
\newcommand{\ex}{\mathbbE}

\def\BibTeX{{\rm B\kern-.05em{\sc i\kern-.025em b}\kern-.08em
    T\kern-.1667em\lower.7ex\hbox{E}\kern-.125emX}}
\begin{document}

\title{Attacks and Defenses for Generative Diffusion Models: A Comprehensive Survey
	\thanks{V. T. Truong, L. B. Dang, and L. B. Le are with INRS, University of Qu\'{e}bec, Montr\'{e}al, QC H5A 1K6, Canada (email: tuan.vu.truong@inrs.ca, ba.luan.dang@inrs.ca, long.le@inrs.ca).}
}

\author{\IEEEauthorblockN{Vu Tuan Truong, Luan Ba Dang, and Long Bao Le}
% \IEEEauthorblockA{
% \textit{INRS-EMT, University of Québec}\\
% Montréal, Québec, Canada \\
% tuan.vu.truong@inrs.ca, long.le@inrs.ca}
}

\maketitle

\begin{abstract}
Diffusion models (DMs) have achieved state-of-the-art performance on various generative tasks such as image synthesis, text-to-image, and text-guided image-to-image generation. However, the more powerful the DMs, the more harmful they potentially are. Recent studies have shown that DMs are prone to a wide range of attacks, including adversarial attacks, membership inference, backdoor injection, and various multi-modal threats. Since numerous pre-trained DMs are published widely on the Internet, potential threats from these attacks are especially detrimental to the society, making DM-related security a worth investigating topic. Therefore, in this paper, we conduct a comprehensive survey on the security aspect of DMs, focusing on various attack and defense methods for DMs. First, we present crucial knowledge of DMs with five main types of DMs, including denoising diffusion probabilistic models, denoising diffusion implicit models, noise conditioned score networks, stochastic differential equations, and multi-modal conditional DMs. We further survey a variety of recent studies investigating different types of attacks that exploit the vulnerabilities of DMs. Then, we thoroughly review potential countermeasures to mitigate each of the presented threats. Finally, we discuss open challenges of DM-related security and envision certain research directions for this crucial topic.
% consisting of four main aspects: (i) different attacks against DMs, (ii) defense methods for DMs, (iii) using DMs to attack other systems, and (iv) DMs as a mean of defense. Each aspect is investigated thoroughly in a well-structured section, in which we categorize the attacks/defenses into different strategies according to state-of-the-art studies. In the last part of the study, we provide a detailed discussion of current limitations and open challenges of DM-related security, while envisioning future research directions for this crucial topic.
\end{abstract}

\begin{IEEEkeywords}
Diffusion models, multi-modal threats, diffusion model security, backdoor attack, membership inference, adversarial attacks.
\end{IEEEkeywords}

\section{Introduction}\label{section:introduction}
In recent years, diffusion models (DMs)\cite{sohl2015deep, ho2020denoising, dhariwal2021diffusion, nichol2021improved, song2020denoising, song2019generative, song2021maximum, song2020score, rombach2022high, song2020improved, vahdat2021score} have showcased remarkable capability in a wide range of generative tasks, setting the new state-of-the-art among other categories of deep generative models such as generative adversarial networks (GANs)\cite{goodfellow2014generative}, variational autoencoders (VAEs)\cite{kingma2014vae, rezende2015variational}, and energy-based models (EBMs)\cite{ngiam2011learning}. In general, DMs consist of two main processes. The forward (diffusion) process progressively adds noise to the original data to gradually diffuse the data distribution into the standard Gaussian distribution. The reverse (generative) process employs a deep neural network, which is often a UNet\cite{ronneberger2015u}, to reverse the diffusion, reconstructing the data from the Gaussian noise. With its impressive potentials, DMs have been investigated in various domains, including computer vision\cite{watson2021learning, nichol2021glide, sinha2021d2c, bao2022analytic, dockhorn2021score, avrahami2022blended, ramesh2022hierarchical, liu2022compositional, daniels2021score, chung2022come, esser2021imagebart, lugmayr2022repaint}, natural language processing (NLP)\cite{austin2021structured, hoogeboom2021argmax, li2022diffusion, savinov2021step, yu2022latent, lin2023text}, audio processing\cite{chen2020wavegrad, popov2021grad}, 3D generation\cite{xu2023dream3d, truong2024text, poole2022dreamfusion, lin2023magic3d, haque2023instruct, luo2021diffusion}, bioinformatics\cite{xu2022geodiff, luo2022antigen}, and time series tasks\cite{tashiro2021csdi, yan2021scoregrad, rasul2020multivariate}.

DMs can be classified into different categories based on their diffusion and generative processes. The first category, inspired by the non-equilibrium thermodynamics theory\cite{sohl2015deep}, comprises of denoising diffusion probabilistic models (DDPMs)\cite{ho2020denoising, sohl2015deep, dhariwal2021diffusion, nichol2021improved}. DDPMs can be viewed as a Markovian Hierarchical VAE, in which the diffusion process is modeled as a Markov chain with multiple consecutive VAEs. Each diffusion step corresponds to the encoding process of a VAE, while each denoising step can be considered a decoding operation of the corresponding VAE. On the other hand, denoising diffusion implicit models (DDIMs), a variant of DDPMs, take a non-Markovian approach that allows the models to skip steps in the denoising process, thus increasing the generating speed with a certain trade-off for quality. Another category of DMs is noise conditioned score networks (NCSNs)\cite{song2019generative, song2020improved, song2021maximum}, in which a neural network is trained based on score matching\cite{hyvarinen2005estimation} to learn the score function (i.e., the gradient of the log likelihood) of the true data distribution. This score function points towards the data space that the training data inhabits. Therefore, by following the score, well-trained NCSNs can generate new samples accordingly to the true data distribution. This process can also be viewed as a denoising process\cite{vincent2011connection}. The final main category, score-based stochastic differential equation (SDE), encapsulates both DDPMs and NCSNs into a generalized form. While the forward process maps data to a noise distribution via an SDE, the reverse process uses the reverse-time SDE\cite{anderson1982reverse} to generate samples from noise. Moreover, cross-attention technique\cite{vaswani2017attention} can be used to constraint the denoising neural network with multi-modal conditions such as text and images, forcing the denoising process to generate results that follow the given conditions. This sparked a wide variety of multi-modal generative tasks such as text-to-image and text-guided image-to-image generation\cite{rombach2022high}.

\begin{table*}[]
\centering
\caption{A comparison between our study and existing surveys regarding DMs and its applications}
\label{table:survey-comparison}
\renewcommand{\arraystretch}{1.2}
\begin{tabular}{|c|c|l|}
\hline
\rowcolor[HTML]{EFEFEF} 
\textbf{Survey} & \multicolumn{1}{c|}{\cellcolor[HTML]{EFEFEF}\textbf{DM-Related Topic}} & \multicolumn{1}{c|}{\cellcolor[HTML]{EFEFEF}\textbf{Key Contents}} \\ \hline
\cite{yang2023diffusion} & DM Improvements & \begin{tabular}[c]{@{}l@{}}Efficient sampling, improved likelihood, DMs for special-structure data, DM-based  applications,\\  connection between DMs and other generative models\end{tabular} \\ \hline
\cite{10419041} & DM Improvements & \begin{tabular}[c]{@{}l@{}}Sampling acceleration, diffusion process design, likelihood optimization, bridging distributions, \\ DM-based general applications\end{tabular} \\ \hline
\cite{10081412} & Computer Vision & \begin{tabular}[c]{@{}l@{}}Computer vision applications, including text-to-image \& image-to-image generation, image \\ super-resolution, editing, segmentation, anomaly detection\end{tabular} \\ \hline
\cite{zou2023diffusion} & NLP & \begin{tabular}[c]{@{}l@{}}DM-based NLP applications such as machine translation, text generation, image captioning, \\ summarization, completion, style transfer, paraphrasing\end{tabular} \\ \hline
\cite{kazerouni2023diffusion} & Medical & \begin{tabular}[c]{@{}l@{}}DMs in medical imaging: Clinical image translation, reconstruction, registration, classification,\\  segmentation, denoising, generation, anomaly detection\end{tabular} \\ \hline
\cite{lin2023diffusion} & Time Series & \begin{tabular}[c]{@{}l@{}}DMs for time-series applications, including time-series forecasting, imputation, generation, \\ and further applications in other domains\end{tabular} \\ \hline
\textbf{Ours} & \textbf{DM-Related Security} & \begin{tabular}[c]{@{}l@{}}Security of DMs, including different types of attacks (e.g., backdoor, membership inference,\\  adversarial), the corresponding countermeasures, and DM-based attacking/defense methods\end{tabular} \\ \hline
\end{tabular}
\end{table*}

Despite these remarkable potentials, DMs are especially vulnerable to various security and privacy risks due to the following reasons: (i) robust DMs are often trained on large-scale data collected from diverse open sources, which might include poisoned or backdoored data; (ii) pre-trained DMs are published widely on open platforms like HuggingFace\footnote{https://huggingface.co/}, making it easier for hackers to spread their manipulated models. For instance, by manipulating the training data and modifying the training objective, attackers can embed a backdoor trigger into DMs to conduct a backdoor attack\cite{10494544, Chou2023CVPR, huang2024personalization, struppek2023rickrolling, chen2023trojdiff, chou2024villandiffusion, wang2023stronger, pan2023trojan, zhai2023text, li2023learnable}. Consequently, once the trigger is fed into the backdoored DM during inference, it will consistently produce a specific result that is designated by the attackers (e.g., a sensitive image or a violent text). Even in a more secure setting in which the attackers cannot modify the DMs' parameters, they can still craft the inputs of eligible DMs to make it generates sensitive contents; this is referred to as adversarial attack\cite{salman2023raising, van2023anti, yu2024step, shan2023glaze, zhang2023robustness, yang2023mma, yang2024sneakyprompt, zhuang2023pilot, gao2023evaluating, zhang2024revealing, liu2023discovering, liu2023riatig, kou2023character, zhang2023generate, liang2023adversarial, liang2023mist, zhu2024watermark}. In terms of privacy, membership inference can be launched to detect whether a particular example was included in the training dataset of DMs. This is especially dangerous when the training data is highly sensitive (e.g., medical images). Furthermore, as DMs are also used in various security applications such as adversarial purification and robustness certification, attacking the DMs integrated in such applications can disable the entire DM-based security systems\cite{li2023change, kang2024diffattack}.

Since DMs are gaining significant attention and various DM-based applications have been used widely by the public, it is undeniable that security of DMs is an important research direction.
However, existing surveys on DMs mostly exploit its developments in terms of architecture improvements, performance, and applications, while the security aspect of DMs is totally neglected.
For instance, the authors in\cite{10419041} surveyed a wide range of algorithm improvements for DMs, including the improvements in sampling acceleration, diffusion process design, likelihood optimization, and bridging  distributions. Besides, they also reviewed various applications of DMs such as image/video generation, medical analysis, text generation, and audio generation. Similarly, the survey\cite{yang2023diffusion} also discussed the applications and development of DMs, with a significant attention on efficient sampling methods and improved likelihood. Furthermore, the authors investigated thoroughly the connection between DMs and other categories of deep generative models like VAEs, GANs, and EBMs. In terms of application-centric surveys, there are multiple surveys study DM-based applications, including computer vision\cite{10081412}, NLP\cite{zou2023diffusion}, medical imaging\cite{kazerouni2023diffusion}, and time-series applications\cite{lin2023diffusion}.

As none of existing surveys investigates the security aspect of DMs, this paper aims to fill the gap by providing a systematic and comprehensive overview of state-of-the-art research studies in this crucial topic. By categorizing different types of DM-targeted attacks and presenting countermeasures for tackling these attacks, we hope this survey provides a helpful guideline for researchers to explore and develop state-of-the-art security methods for DMs. The contribution of the paper can be summarized as follows:
\begin{itemize}
    \item We provide readers with necessary background knowledge of different types of DMs, including DDPM, DDIM, NCSN, SDE, and multi-modal conditional DMs. We demonstrate how different categories of DMs relate to each other under a consistent diffusion principle.
    \item We investigate a wide range of attacks on DMs, categorized into three main groups, including backdoor attacks, membership inference attacks (MIAs), and adversarial attacks. Each attack is categorized further into sub-groups based on the corresponding methods/ applications.
    \item We survey various countermeasures for DM-targeted attacks based on state-of-the-art research studies in that field. 
    \item We discuss multiple open challenges in this topic and envision some interesting research directions to improve the security aspect of DMs and DM-based applications.
\end{itemize}

Table~\ref{table:survey-comparison} presents a comparison between our work and existing DM-related surveys, emphasizing our contributions. The rest of this paper is presented as follows. Section~\ref{section:background} provides preliminaries of different types of DMs and background knowledge of DM security. Section~\ref{section:attack-DM} surveys state-of-the-art methods for attacking DMs and DM-based systems. Then, different countermeasures for the presented attacks are discussed in Section~\ref{section:defense-DM}. Section~\ref{section:open-challenges} discusses various open challenges and future research directions on this field, while Section~\ref{section:conclusion} concludes our survey.

\section{Background Knowledge}\label{section:background}
\begin{figure}[!t]
    \centering
    \begin{subfigure}[b]{\columnwidth}
        \centering
        \includegraphics[width=\textwidth]{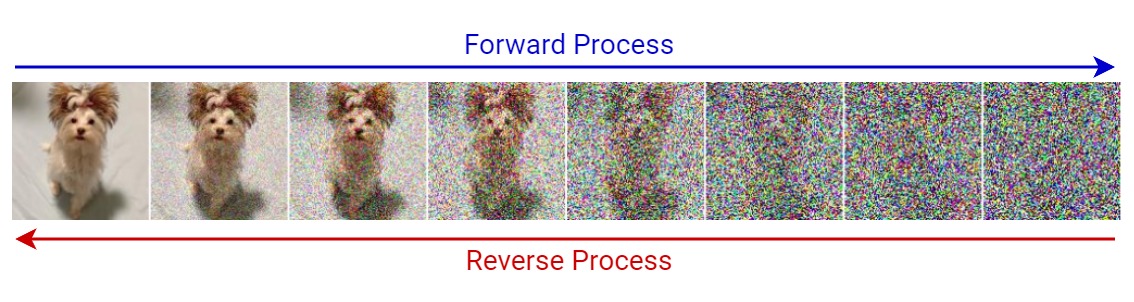}
        \caption{The general processes of typical DMs.}
        \label{subfig:general}
    \end{subfigure}
    \vskip\baselineskip
    \begin{subfigure}[b]{\columnwidth}
        \centering
        \includegraphics[width=\textwidth]{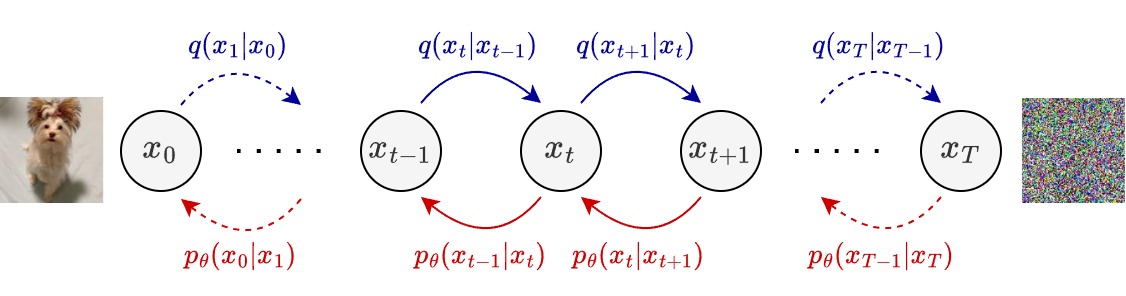}
        \caption{Denoising Diffusion Probabilistic Models.}
        \label{subfig:ddpm}
    \end{subfigure}
    \vskip\baselineskip
    \begin{subfigure}[b]{\columnwidth}
        \centering
        \includegraphics[width=\textwidth]{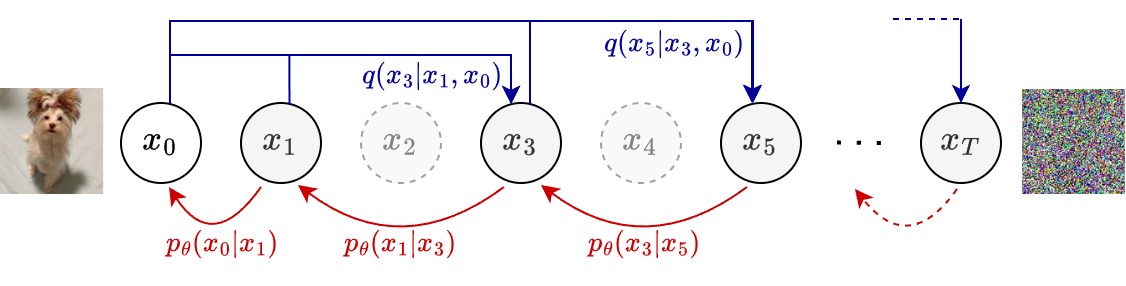}
        \caption{Denoising Diffusion Implicit Models. In this example, the DDIM skips one step after every single step, thus doubling the speed.}
        \label{subfig:ddim}
    \end{subfigure}
    \vskip\baselineskip
    \begin{subfigure}[b]{\columnwidth}
        \centering
        \includegraphics[width=\textwidth]{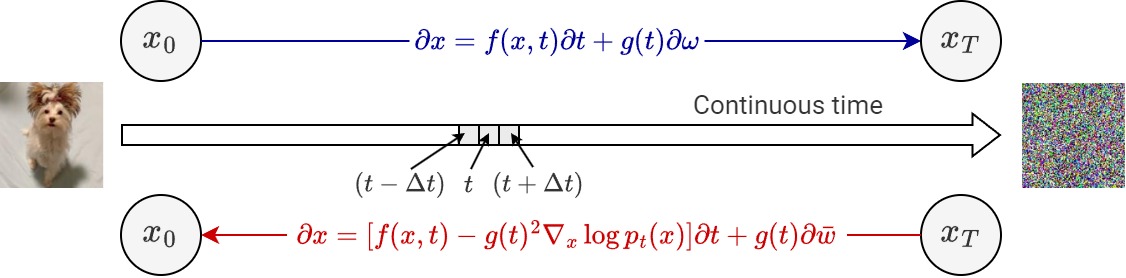}
        \caption{Score Stochastic Differential Equation.}
        \label{subfig:sde}
    \end{subfigure}
    \vskip\baselineskip
    \begin{subfigure}[b]{0.85\columnwidth}
        \centering
        \includegraphics[width=\textwidth]{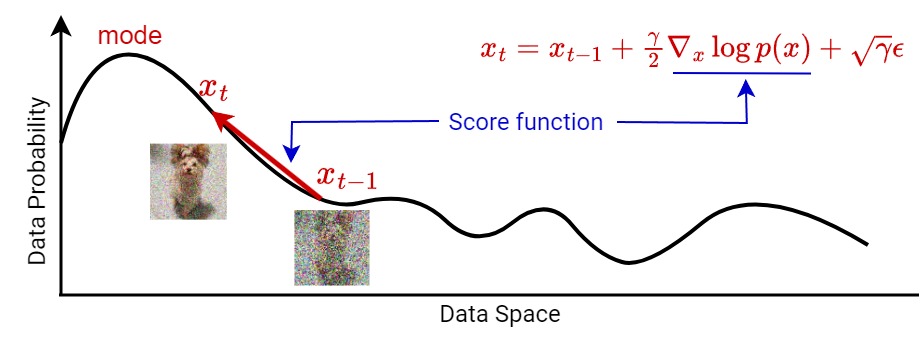}
        \caption{Noise Conditioned Score Networks.}
        \label{subfig:ncsn}
    \end{subfigure}
    \caption{Diffusion models viewed from different perspectives according to four main categories, including DDPMs, DDIMs, NCSNs, and SDE.}
    \label{fig:DM}
\end{figure}

DM is a type of deep generative models that learn to generate samples from random noise via two processes, which are the forward diffusion process and the reverse process. These processes are illustrated in Fig.~\ref{subfig:general}. Given a training data sample such as an image, the diffusion process progressively adds noise into the image. After $T$ noising steps, the original training data sample at step 0 is completely destroyed, resulting in a standard Gaussian noise at step $T$. Then, in the denoising process, a deep neural network, which is often a UNet\cite{ronneberger2015u}, is trained to predict and remove the noise gradually from step $T$ back to step $0$, thus reconstructing the original image. After finishing the training process on a large-scale dataset, the UNet can generate new samples from any arbitrary Gaussian noise, in which the generated samples lie on the space of the original training data. 

In practice, the denoising process can be viewed from different perspectives according to three main categories of DMs, which are DDPM, NCSN, and score-based SDE. 
% DDPMs model both the forward and reverse processes as Markov chains; NCSNs follow the data's score function via score matching for denoising; and score-based SDE treats the denoising process as a reverse-time SDE problem\cite{anderson1982reverse}. 
Although different categories of DMs result in varied generation performance, they all can be interpreted via the presented noising and denoising processes. Besides the above three categories, this section also presents two other important variants of DMs. The first one is DDIM, a faster and non-Markovian version of DDPM, while the second one is multi-modal DMs, which are DMs with additional multi-modal conditions constrained by the UNet's cross-attention layers (e.g., stable/latent diffusion).

\subsection{Denoising Diffusion Probabilistic Models (DDPMs)}\label{subsection:DDPM}
DDPMs model both the forward and reverse processes as Markov chains, illustrated in Fig.~\ref{subfig:ddpm}. This means that the result of each step only depends on its previous step, and there is a stochastic factor during any step transition.

\subsubsection{Forward Process}
The forward process of DDPMs is modelled as a Markov chain, which progressively adds noise to the clear data sample $x_0$ in a total of $T$ steps to generate a sequence of noisy distributions $x_1, x_2,...,x_T$. The transition between two consecutive diffusion steps is defined as:
\begin{equation}
\label{equation1}
    q(x_t|x_{t-1}) = \mathcal{N}(x_t;\sqrt{\alpha_t}x_{t-1}, (1-\alpha_t)\mathbf{I}),
\end{equation}
where $\alpha_t \in (0,1)$ is the noise schedule that determines the noise scale at each diffusion step $t$. While $\alpha_t$ can be either a fixed or a learnable hyperparameter, it must be chosen such that the final distribution $x_T$ becomes a standard Gaussian, i.e., $p(x_T)=\mathcal{N}(x_T; 0, \mathbf{I})$. In general, $\alpha_t$ often decreases over time, which means that the noise increases gradually from step $0$ to step $T$.
Then, we can apply the reparameterization trick on (\ref{equation1}) to sample $x_t$ from $x_{t-1}$ with a sampled noise $\epsilon \sim \mathcal{N}(0,\mathbf{I})$:
\begin{equation}
\label{equation2}
    x_t = \sqrt{\alpha_t}x_{t-1} + \sqrt{1-\alpha_t}\epsilon.
\end{equation}

By repeating the reparameterization trick recursively based on (\ref{equation2}), the choice of (\ref{equation1}) leads to an important property that allows the direct sampling of $x_t$ from the clear data $x_0$:
\begin{equation}
\label{equation3}
    q(x_t|x_0) = \mathcal{N}(x_t; \sqrt{\Bar{\alpha}_t}x_0, (1-\Bar{\alpha}_t)\mathbf{I})
\end{equation}
\begin{equation}
\label{add_noise_ddpm}
    x_t = \sqrt{\Bar{\alpha}_t}x_0 + \sqrt{1-\Bar{\alpha}_t}\epsilon_0,
\end{equation}
where $\Bar{\alpha}_t = \prod_{i=1}^{t} \alpha_{i}$ and $\epsilon_0 \sim \mathcal{N}(0,\mathbf{I})$.
% Thanks to this property, we do not have to store the variable of every $x_t$ when training the UNet, which costs significant storage when $T$ is large.

\subsubsection{Reverse Process} The reverse process uses deep neural networks parameterized by $\theta$ with the following denoising transitions:
\begin{equation}
\label{equation5}
    p_\theta(x_{t-1}|x_t) = \mathcal{N}(x_{t-1}; \mu_\theta(x_t,t), \Sigma_\theta(x_t,t)),
\end{equation}
where $\mu_\theta(x_t,t)$ and $\Sigma_\theta(x_t,t)$ are the mean and variance parameterized by a deep neural network. 
The neural network approximates the data distribution via the log likelihood:
\begin{equation}
\label{equation6}
    \log p_\theta(x_0) = \log \int p_\theta(x_{0:T})dx_{1:T},
\end{equation}
with $p_\theta(x_{0:T})=p(x_T)\prod_{t=1}^Tp_\theta(x_{t-1}|x_t)$.

\begin{table*}[]
\centering
\caption{A summary of training objectives and sampling processes for each category of DMs.}
\label{table:dm-comparison}
\renewcommand{\arraystretch}{2.2} 
\begin{tabular}{|c|l|l|}
\hline
\textbf{Category} & \multicolumn{1}{c|}{\textbf{Training Objective}} & \multicolumn{1}{c|}{\textbf{Sampling/Inference}} \\ [0.3ex]
 \hline
DDPM & \multirow{2}{*}{$\mathcal{L}^\text{ddpm}_\text{simple}= \mathcal{L}^\text{ddim} =\ex_{x_0,\epsilon_0} \left[ \left\| \epsilon_0 - \epsilon_\theta(x_t,t) \right\|^2_2 \right]$} & $x_{t-1} = \frac{1}{\sqrt{\alpha_t}}(x_t - \frac{1-\alpha_t}{\sqrt{1-\Bar{\alpha}_t}}\epsilon_\theta(x_t,t)) + \sqrt{\widetilde{\beta}_t}\epsilon$ \\ [0.6ex] \cline{1-1} \cline{3-3} 
DDIM &  & $x_{t-1} = \sqrt{\Bar{\alpha}_{t-1}} \left( \frac{x_t-\sqrt{1-\Bar{\alpha}_t}\epsilon_0^{(\theta)}(x_t,t)}{\sqrt{\Bar{\alpha}_t}} \right) + \sqrt{1-\Bar{\alpha}_{t-1}}\epsilon_0^{(\theta)}(x_t,t)$ \\ [0.6ex] \hline
NCSN & $\mathcal{L}^\text{ncsn} = \frac{1}{2T} \sum_{t=1}^T \lambda(\sigma_t) \ex_{p,\widetilde{x}} \left\| s_\theta(\widetilde{x},\sigma_t) + \frac{\widetilde{x}-x}{\sigma_t^2} \right\|^2_2$ & $\widetilde{x}_t = \widetilde{x}_{t-1}+ \frac{\gamma_t}{2}s_\theta(\widetilde{x}_{t-1},\sigma_t) +\sqrt{\gamma_t}\epsilon$ \\ [1.2ex] \hline
SDE & $\mathcal{L}^\text{sde} = \ex_{t,p_0,p_t} \left[ \lambda(t)\left\| s_\theta(x_t,t) - \nabla_{x_t}\log {p_t}(x_t|x_0) \right\|^2_2 \right]$ & $\partial x = [f(x,t) - g(t)^2\nabla_x\log p_t(x)]\partial t + g(t)\partial \Bar{w}$ \\ [0.6ex] \hline
\end{tabular}
\end{table*}

\subsubsection{Training and Sampling} Using Jensen's inequality, the problem of maximizing the log likelihood in (\ref{equation6}) can be interpreted as maximizing its evidence lower bound (ELBO), resulting in the following training objective:
\begin{equation}
\label{equation7}
\begin{split}
    \mathcal{L}^\text{ddpm} = \ex_q [ -\log p_\theta(x_0|x_1) + D_\text{KL}(q(x_T|x_0)||p(x_T)) \\ + \sum_{t=2}^{T}D_\text{KL}(q(x_{t-1}|x_t,x_0)||p_\theta(x_{t-1}|x_t)) ],
\end{split}
\end{equation}
where $D_\text{KL}$ denotes the KL divergence between two distributions. It can be seen that the second term in (\ref{equation7}) can be ignored since it has no trainable parameter. Although the first term can be optimized using Monte Carlo estimate, it is just a single denoising step, which is totally dominated by the final term which consists of $T-1$ steps; thus the first term can be ignored in practice. Consequently, only the final summation term remains in the training objective. To optimize this final term, we respectively analyze $q(x_{t-1}|x_t,x_0)$ and $p_\theta(x_{t-1}|x_t)$. By applying the Bayes rule, we obtain:
\begin{equation}
\label{equation8}
    q(x_{t-1}|x_t,x_0) = \frac{q(x_t|x_{t-1},x_0)q(x_{t-1}|x_0)}{q(x_{t}|x_0)}.
\end{equation}

We can derive both $q(x_{t-1}|x_0)$ and $q(x_{t}|x_0)$ from (\ref{equation3}), while $q(x_t|x_{t-1},x_0)$ is derived from (\ref{equation1}). As a result, the above posterior can be expressed as a Gaussian distribution $q(x_{t-1}|x_t,x_0) = \mathcal{N}(x_{t-1};\mu_q(x_t,\epsilon_0),\Sigma_q(t))$, with:
\begin{equation}
\label{equation9}
    \mu_q(x_t,\epsilon_0) = \frac{1}{\sqrt{\alpha_t}}x_t - \frac{1-\alpha_t}{\sqrt{1-\Bar{\alpha}_t}\sqrt{\alpha_t}}\epsilon_0.
\end{equation}
\begin{equation}
\label{equation10}
    \Sigma_q(t) = \sigma_q^2(t) = \frac{(1-\alpha_t)(1-\Bar{\alpha}_{t-1})}{1-\Bar{\alpha}_{t}}
\end{equation}

In terms of $p_\theta(x_{t-1}|x_t)$ (\ref{equation5}), Ho et al.~\cite{ho2020denoising} opt to fix the variance $\Sigma_\theta(x_t,t)$ to $\Sigma_q(t)$, while the mean $\mu_\theta(x_t, t)$ is chosen with the same form of $\mu_q(x_t,\epsilon_0)$ but the noise $\epsilon_0$ is replaced by the neural network's predicted noise $\epsilon_\theta$:
\begin{equation}
\label{equation11}
    \mu_\theta(x_t, t) = \frac{1}{\sqrt{\alpha_t}}x_t - \frac{1-\alpha_t}{\sqrt{1-\Bar{\alpha}_t}\sqrt{\alpha_t}}\epsilon_\theta(x_t,t),
\end{equation}
where $\epsilon_\theta(x_t,t)$ is the noise predicted by the neural network.

It is showed that the KL divergence between two Gaussian distributions with the same variance can be reduced to the difference between their means\cite{luo2022understanding}. Thus, the training objective of DDPMs becomes:
\begin{align}
\mathcal{L}^\text{ddpm} &= \ex_q \left[ D_\text{KL}(q(x_{t-1}|x_t,x_0)||p_\theta(x_{t-1}|x_t)) \right] \\
&= \ex_{x_0,\epsilon_0} \left[ \frac{1}{2\sigma_q^2(t)} \left\| \mu_q(x_t,\epsilon_0) -\mu_\theta(x_t, t) \right\|^2_2 \right] \\
&= \ex_{x_0,\epsilon_0} \left[ \frac{1}{2\sigma_q^2(t)} \frac{(1-\alpha_t)^2}{(1-\Bar{\alpha}_t)\alpha_t} \left\| \epsilon_0 - \epsilon_\theta(x_t,t) \right\|^2_2 \right]
\end{align}

Here, $x_t$ is a function of $x_0$ and $\epsilon_0$, sampled from (\ref{equation2}). Ho et al.\cite{ho2020denoising} empirically showed that simplifying the objective function to the following form leads to a better sample quality:
\begin{equation}
\label{equation15}
    \mathcal{L}^\text{ddpm}_\text{simple} = \ex_{x_0,\epsilon_0} \left[ \left\| \epsilon_0 - \epsilon_\theta(x_t,t) \right\|^2_2 \right]
\end{equation}

Intuitively, the neural network is trained to predict the source noise corresponding to each time step. Then, the neural network can be used for sampling to generate samples:
\begin{equation}
\label{equation16}
    x_{t-1} = \frac{1}{\sqrt{\alpha_t}} \left( x_t - \frac{1-\alpha_t}{\sqrt{1-\Bar{\alpha}_t}}\epsilon_\theta(x_t,t) \right) + \sqrt{\widetilde{\beta}_t}\epsilon,
\end{equation}
where $\widetilde{\beta}_t = \frac{(1-\alpha_t)(1-\Bar{\alpha}_{t-1})}{1-\Bar{\alpha}_{t}}$, which is the variance in (\ref{equation10}).

\subsection{Denoising Diffusion Implicit Models (DDIMs)}
In DDPMs, the number of denoising steps $T$ is often chosen to be large (e.g., 1000 steps in\cite{ho2020denoising}) to make the reverse process close to a Gaussian distribution\cite{sohl2015deep}. With a low value of $T$, the generation results degrade significantly since the reverse process modeled with Gaussian distributions is no more a good approximation. However, such a large $T$ leads to a low sampling/generating speed since all the steps must be performed sequentially due to the Markov chain's property.
In DDIMs, the inference process is designed to be non-Markovian so we can skip steps in the denoising process, resulting in a faster generating speed with a certain trade-off in terms of generating quality. It is described in Fig.~\ref{subfig:ddim}.

\subsubsection{Forward Process} while DDPMs first choose the diffusion transition $q(x_t|x_{t-1})$ (\ref{equation1}), then use it to derive the posterior $q(x_{t-1}|x_t,x_0)$ (\ref{equation8}) via the Bayes rule, DDIMs take an opposite approach by choosing this posterior first:
\begin{equation}
\label{equation17}
    q_\sigma(x_{t-1}|x_t,x_0) = \mathcal{N}(x_{t-1}; \mu_\sigma(x_t,x_0), \sigma^2_t \mathbf{I}),
\end{equation}
where $\sigma = [\sigma_1,...,\sigma_T]$ is a vector of positive coefficients controlling the stochastic magnitude of the forward process, and the mean $\mu_\sigma(x_t,x_0)$ is chosen as follows:
\begin{equation}
\label{equation18}
    \mu_\sigma(x_t,x_0) = \sqrt{\Bar{\alpha}_{t-1}}x_0 + \sqrt{1-\Bar{\alpha}_{t-1}-\sigma^2_t} \frac{x_t-\sqrt{\Bar{\alpha}_t}x_0}{\sqrt{1-\Bar{\alpha}_t}}.
\end{equation}
The intentional choice of this mean function is to offer the following desirable property: 
\begin{equation}
\label{equation19}
    q_\sigma(x_t|x_0) = \mathcal{N}(x_t; \sqrt{\Bar{\alpha}_t}x_0, (1-\Bar{\alpha}_t)\mathbf{I}),
\end{equation}
which is proved in\cite{song2020denoising}. This property is similar to the one presented in DDPMs (\ref{equation3}). With this property, the forward transition of DDIMs can be derived based on the Bayes rule:
\begin{equation}
    q_\sigma(x_t|x_{t-1},x_0) = \frac{q_\sigma(x_{t-1}|x_t,x_0)q_\sigma(x_t|x_0)}{q_\sigma(x_{t-1}|x_0)}.
\end{equation}

It can be seen that each diffusion step $t$ of DDIMs depends on both $x_{t-1}$ and $x_0$, making it non-Markovian.

\subsubsection{Reverse Process} By recursively applying the reparameterization trick on (\ref{equation19}), we have that:
\begin{equation}
\label{equation21}
    x_t = \sqrt{\Bar{\alpha}_t}x_0 + \sqrt{1-\Bar{\alpha}_t}\epsilon_0.
\end{equation}

We can train a neural network to predict the source noise $\epsilon_0$, thus predicting the corresponding $x_0$ from (\ref{equation21}):
\begin{equation}
    x_0^{(\theta)}(x_t,t) = \frac{x_t-\sqrt{1-\Bar{\alpha}_t}\epsilon_0^{(\theta)}(x_t,t)}{\sqrt{\Bar{\alpha}_t}},
\end{equation}
where $\epsilon_0^{(\theta)}(x_t,t)$ and $x_0^{(\theta)}(x_t,t)$ are the source noise $\epsilon_0$ and the denoised observation $x_0$, respectively, predicted by the neural network with the parameter $\theta$.

Consequently, the reverse process is defined by leveraging the knowledge of $q_\sigma(x_{t-1}|x_t,x_0)$ in (\ref{equation17}), with $x_0$ replaced by the predicted $x_0^{(\theta)}(x_t,t)$:
\begin{align}
    p_\theta(x_{t-1}|x_t) &= q_\sigma(x_{t-1}|x_t,x_0^{(\theta)}(x_t,t)) \\
    &= \mathcal{N}(x_{t-1}; \mu_\sigma(x_t,x_0^{(\theta)}), \sigma^2_t \mathbf{I}), \label{equation24}
\end{align}
where the mean $\mu_\sigma(\cdot)$ is defined in (\ref{equation18}).

\subsubsection{Training and Sampling} The neural network $\theta$ is also optimized by maximizing the log likelihood presented in (\ref{equation6}), which leads to the following ELBO objective:
\begin{equation}
\label{equation25}
\begin{split}
    \mathcal{L}^\text{ddim}_\sigma = \ex_{q_\sigma} [D_\text{KL}(q_\sigma(x_T|x_0)||p(x_T))  -\log p_\theta(x_0|x_1) \\ + \sum_{t=2}^{T}D_\text{KL}(q_\sigma(x_{t-1}|x_t,x_0)||p_\theta(x_{t-1}|x_t)) ].
\end{split}
\end{equation}

Song et al.~\cite{song2020denoising} proved that the optimal solution of $\mathcal{L}^\text{ddim}_\sigma$ is the same as that of $\mathcal{L}^\text{ddpm}_\text{simple}$ in (\ref{equation15}).

In  terms of sampling, $x_{t-1}$ can be sampled from $x_t$ by applying the reparameterization trick on (\ref{equation24}):
\begin{equation}
\label{equation26}
\begin{split}
    x_{t-1} = \sqrt{\Bar{\alpha}_{t-1}} \left( \frac{x_t-\sqrt{1-\Bar{\alpha}_t}\epsilon_0^{(\theta)}(x_t,t)}{\sqrt{\Bar{\alpha}_t}} \right) \\
    + \sqrt{1-\Bar{\alpha}_{t-1}-\sigma_t^2}\epsilon_0^{(\theta)}(x_t,t) + \sigma_t\epsilon_t,
\end{split}
\end{equation}
where $\epsilon_t \sim \mathcal{N}(\mathbf{0},\mathbf{I})$.
Especially, (\ref{equation26}) is equivalent to (\ref{equation16}) if we choose $\sigma_t = \widetilde{\beta}_t$, with $\widetilde{\beta}_t$ defined below the equation (\ref{equation16}). In this case, it becomes a DDPM. On the other hand, it becomes a DDIM if we choose $\sigma_t = 0$. This leads to a deterministic forward process as no noise is added in each step.

Since the generative process is now non-Markovian, we can skip steps to accelerate the generating speed. For example, DDIMs can generate 2 times faster than DDPMs if the sequence of time steps $\mathcal{T}_2 = [1, 3, 5, 7,..., T]$ is used during inference, or 4 times higher than DDPMs with $\mathcal{T}_4 = [1, 5, 9, 13,..., T]$. In other words, we can train DDIMs with an arbitrary large number of time steps $T$, but only sample from some of them (e.g., using $\mathcal{T}_2$ or $\mathcal{T}_4$ for sampling) to speed up the generative process. Notably, it is known that DDPMs are trained based on $\mathcal{T}_1 = [1, 2, 3,..., T]$ . This means that we can take advantages of pretrained DDPMs for DDIMs\cite{song2020denoising}.

\subsection{Noise Conditioned Score Networks (NCSNs)}
Unlike DDPMs and DDIMs which generate samples by predicting and removing noise, NCSNs take another approach by following the score function of the training data to generate samples. The score function of a data density $p(x)$ is defined as the gradient of the log probability density $\nabla_x\log p(x)$. Essentially, the score function indicates the direction in the data space that one needs to move in order to maximize the likelihood of the data $x$. The key idea of NCSNs bases on this property, demonstrated in Fig.~\ref{subfig:ncsn}. Starting from an arbitrary point in the data space (i.e., a random noise), NCSNs are trained to iteratively follow the direction of the score function to move towards the high-density space that the data $x$ inhabits, thereby generating a new sample when reaching a mode of the data distribution. This is referred to as Langevin dynamics, a method originated from physics. Specifically, each step of Langevin dynamics is computed as follows:
\begin{equation}
\label{equation:langevin}
    x_t = x_{t-1} + \frac{\gamma}{2}\nabla_x \log p(x) + \sqrt{\gamma}\epsilon,
\end{equation}
where $t \in \{0,1,...,T\}$, $x_0$ is randomly sampled from a prior distribution, $\gamma$ controls the scale of the update in each step, $\epsilon \sim \mathcal{N}(\mathbf{0},\mathbf{I})$ is a noise that is added to make the generated samples more diverse and stochastic instead of deterministically collapsing onto a local minimum.

Sampling from (\ref{equation:langevin}) only requires the score function. Hence, a neural network parameterized by $\theta$ can be trained to approximate the score function such that $s_\theta(x) \approx \nabla_x \log p(x)$. There are different techniques that can be used to train the neural network, including score matching\cite{hyvarinen2005estimation}, sliced score matching\cite{song2020sliced}, and denoising score matching\cite{vincent2011connection}. For instance, using score matching leads to the following training objective:
\begin{equation}
    \mathcal{L}^\text{sm} = \ex_p \left\| s_\theta(x) - \nabla_x\log p(x) \right\|^2_2.
\end{equation}

However, there are several problems when training with this objective\cite{song2019generative}. By taking the expectation over all examples $p(x)$, rarely-seen examples in low-density space would be dominated by the data in high-density regions. Consequently, score estimation becomes unreliable in low density regions. Another problem is about the manifold hypothesis. Assuming that the data is RGB images. The ambient space that consists of all possible RGB images of size $H\times W$ is obviously very large (with $2^{H\times W \times 3 \times 8}$ possible images), while the training data often lies on only a low dimensional manifold (e.g., car images). As a result, the sampling results would always converge to the low dimensional manifold, while other points outside of this manifold would have probability zero, making the log in the score function ill-defined.

Song et al.\cite{song2019generative} proposed solving these problems by adding a multi-scale Gaussian noise into the data. This noise could push the sampled points to be outside of the low dimensional manifold, solving the manifold hypothesis problem. Moreover, it also increases the regions covered by the modes of the training data, thus covering better low-density regions, mitigating the issue of low data density. Using the noise distribution $q_\sigma(\widetilde{x}|x) = \mathcal{N}(x,\sigma^2_t\mathbf{I})$, the score function can be computed as $\nabla_{\widetilde{x}} \log q_\sigma(\widetilde{x}|x)=-\frac{\widetilde{x}-x}{\sigma^2}$, where $\widetilde{x}$ is a noise-perturbed version of $x$, and $\{ \sigma_t \}^T_{t=1}$ is a sequence of decreasing noise levels. This results in the following objective:

\begin{equation}
    \mathcal{L}^\text{ncsn} = \frac{1}{2T} \sum_{t=1}^T \lambda(\sigma_t) \ex_{p,\widetilde{x}} \left\| s_\theta(\widetilde{x},\sigma_t) + \frac{\widetilde{x}-x}{\sigma_t^2} \right\|^2_2,
\end{equation}
where $\lambda(\cdot)$ is a weighting function, chosen as $\lambda(\sigma)=\sigma^2$ \cite{song2019generative}.

After training the neural network $s_\theta$, it can be used by an annealed version of Langevin dynamics to generate samples:
\begin{equation}
    \widetilde{x}_t = \widetilde{x}_{t-1}+ \frac{\gamma_t}{2}s_\theta(\widetilde{x}_{t-1},\sigma_t) +\sqrt{\gamma_t}\epsilon,
\end{equation}
where $\gamma_t = \mu(\frac{\sigma_t}{\sigma_T})^2$ anneals/scales down the update over time. In practice, $\mu$ is empirically selected between $5\cdot 10^{-6}$ and $5\cdot 10^{-5}$, while $\sigma$ starts from $\sigma_1=1$ to $\sigma_T=0.01$.

Intuitively, score-based generative models like NCSNs naturally link to denoising-based models (e.g., DDPMs and DDIMs) in the sense that: both (i) \textit{denoising} and (ii) \textit{following the score function of data} enable moving in directions that maximize the log probability of the data, starting from a random noise in the data space.

\subsection{Score Stochastic Differential Equation (SDE)}
While DDPMs, DDIMs, and NCSNs operate in a discrete-time scheme with $T$ iterative steps, score SDE generalizes these models to a continuous diffusion process. As illustrated in Fig.~\ref{subfig:sde}, the forward process uses an SDE to diffuse a data example into random noise. The reverse process employs a deep neural network to approximate the reverse-time SDE\cite{anderson1982reverse}, thus generating samples based on a numerical SDE solver such as predictor-corrector samplers and probability flow ordinary differential equation (ODE). In particular, the SDE for the diffusion process is defined as follows:
\begin{equation}
    \partial x = f(x,t)\partial t + g(t)\partial\omega,
\end{equation}
where $t \in [0,T]$ is a continuous time variable, $f(\cdot,t)$ is a function computing the drift coefficient of $x(t)$, $g(\cdot)$ computes the diffusion coefficient, and $w$ is the Brownian motion. The drift term represents the deterministic part of the SDE, responsible for data destruction, while the diffusion coefficient controls the scale of noise from the stochastic part $\partial\omega$. 
On the other hand, the reverse-time SDE has the following form:
\begin{equation}
    \label{equation32}
    \partial x = [f(x,t) - g(t)^2\nabla_x\log p_t(x)]\partial t + g(t)\partial \Bar{w},
\end{equation}
where $\Bar{w}$ is the Brownian motion when the time is reversed back from $T$ to $0$. Essentially, the reverse process starts from a random noise $x_T$, then reverses the time to $x_0$ to generate samples. From a denoising perspective, the term $g(t)^2\nabla_x\log p_t(x)$ can be viewed as the noise that one needs to remove to reconstruct clear samples. Similar to NCSNs, the score function $\nabla_x\log p_t(x)$ here can also be approximated by a deep neural network $s_\theta(x,t)$ trained with denoising score matching, resulting in the following objective:
\begin{equation}
    \mathcal{L}^\text{sde} = \ex_{t,p_0,p_t} \left[ \lambda(t)\left\| s_\theta(x_t,t) - \nabla_{x_t}\log {p_t}(x_t|x_0) \right\|^2_2 \right],
\end{equation}
where $\lambda(t)$ is a weighting function, while the time variable $t$ is sampled over the distribution $\mathcal{U}([0,T])$. Besides denoising score matching\cite{vincent2011connection}, other methods such as sliced score matching\cite{song2020sliced} and finite-difference score matching\cite{pang2020efficient} are also applicable in this case.

\subsection{Multi-Modal Conditional Diffusion Models}\label{subsection:multimodal-DM}
When a DM is trained on a particular dataset, it can generate high-quality samples that lies on the space of the training data. However, the generating results are uncontrollable as it depends on the random noise $x_T$ and other stochastic factors like $\epsilon_t$ during the sampling process. One way to mitigate this issue is to add a condition $c$ at each transition step of DMs:
\begin{equation}
    p(x_{0:T}|c) = p(x_T) \prod_{t=1}^T p_\theta(x_{t-1}|x_t,c).
\end{equation}

Here, the condition $c$ could be in any modality such as text, image, audio, semantic map, or their representation in the latent space. For instance, in terms of DM-based image generation, once might constraint the DM with a text condition $c$ to force it to generate images that follow the textual description provided in $c$. This is referred to as text-to-image generation, a popular application of stable diffusion\cite{rombach2022high}. In DDPMs and DDIMs, the neural networks are trained to predict $\epsilon_\theta(x_t,t,c)$ that approximates the source noise $\epsilon_0$, while the score function $\nabla\log p(x_t|c)$ in NCSNs and SDE models is predicted by $s_\theta(x_t,t,c)$. To constraint the neural networks with such conditions, Rombach et al.\cite{rombach2022high} proposed adding cross-attention layers into the UNet backbone, in which the condition $c$ acts as the key and value of these attention layers, while the query is the data flow forwarded along the UNet architecture. As a result, the supplemented conditions guide the neural network to produce results that follow the added constraints. This turns DMs into a flexible multi-modal generator, enabling a wide variety of generative tasks such as text-to-image, text-guided image-to-image, and text-to-3D generation.

\begin{figure*}[h!]
	\centering
	\includegraphics[scale=0.23]{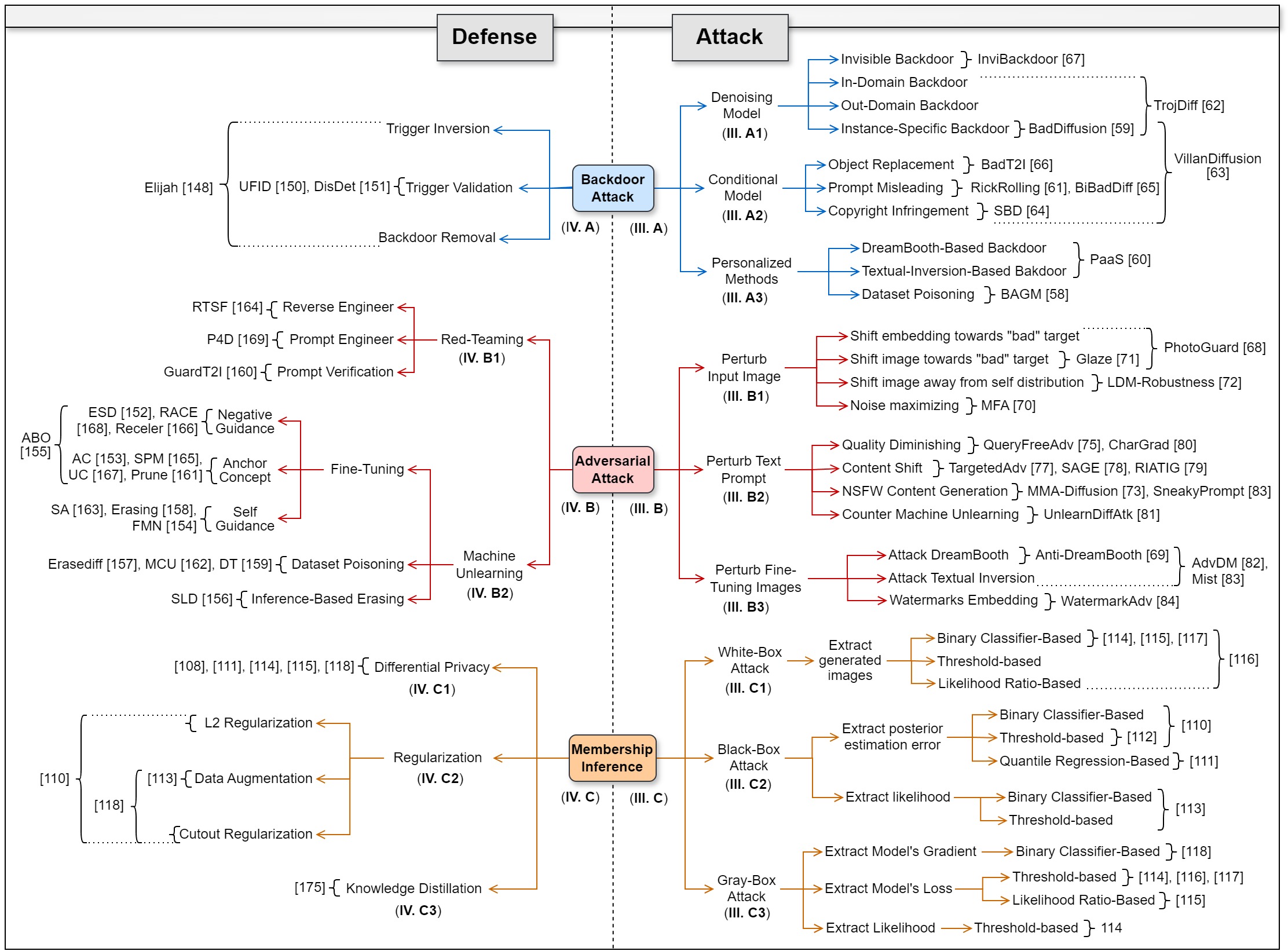}
	\caption{A summary of our survey on attack and defense methods for DMs.}
	\label{figure:overall-survey}
\end{figure*}

\subsection{Preliminary of Diffusion Model Attacks}
\subsubsection{Backdoor Attacks}
In backdoor attacks\cite{li2022backdoor}, the attackers modify the training data and objective function to embed a backdoor trigger into machine learning (ML) models. During inference, if the trigger is activated as input, the models will produce abnormal results driven by the attackers. On the other hand, the models still behave normally on benign samples. This stealthy property of backdoor attacks makes it challenging to be recognized by average users and even detection systems. 
In the context of DMs, backdoor attacks cause DMs to generate a particular image (designated by the attackers) when the trigger is used; this image is called backdoor target.

The purposes of backdoor attack on DMs are diverse. For instance, since many reputable organizations are offering DMs as a service, attackers might inject poisoned data to the training process to backdoor their DMs with sensitive/violent contents as backdoor targets. Then, when the backdoored DMs are used widely by the public, the attackers could publish the backdoor triggers so that everyone can use the triggers to generate the designated sensitive contents, posing a severe threat to both the community and the organizations in terms of reputation and legality. On the other hand, attackers can train their own backdoored DMs, then publish them on open-source platforms like HuggingFace to harm users who download, use, or fine-tune the backdoored models. This risk is even more harmful for downstream applications built upon the backdoored DMs. In this case, a fault caused by the backdoored DMs can make the entire system collapses. Backdoored DMs can be used for advertisement, generating fake news, copyrights. For instance, attackers could manipulate the language model of DMs, making the DMs always generates images related to Starbucks when there is a trigger word ``coffee" in the text prompt.

\subsubsection{Adversarial Attacks}
Adversarial attacks on traditional models likes image classifiers have been investigated extensively in existing studies\cite{madry2017towards, huang2017adversarial}. In general, an adversarial attack is conducted by adding a small perturbation into the input of a neural network, causing the model to produce incorrect answers\cite{8294186}. The perturbation is often learnable, while its scale is constrained to be small enough to ensure the stealthiness of the attack. For instance, regarding object detection for autopilot, an attacker might conduct an adversarial attack by learning a tiny perturbation to mislead the object-detection model. While this perturbation is too small to be recognized by human observers, it might make the autopilot system misunderstands a stop sign as a slow sign. As a result, the autopilot system decides to keep moving, potentially causing severe traffic accidents.

In terms of adversarial attacks on generative DMs, a learned perturbation added into the model's input could cause it to generate chaotic content. This type of attack can be dangerous in various situations. First, an attacker can deploy adversarial attacks to make a well-known public DM generate Not-Safe-For-Work (NSFW) content like racism, horror, politics, pornography, or violence. This not only impacts negatively on the reputation of the organization who published the attacked DM, but also poses a threat to the society in terms of ethic and morality. Second, the output of an DM can be used as input for other downstream tasks in a large-scale system. If the DM is attacked such that it produces manipulated results, it might lead to the failure of the entire system. The effect of an adversarial attack can be diverse, including image quality degradation, content shifting, sensitive content generation, and object distortion/elimination.

On the other hand, adversarial attacks on DMs can also be utilized for positive purposes like copyrights protection, watermark embedding, and anti-personalization. For instance, artists can add a learned perturbation into their arts to hinder DMs from extracting the arts' features. Consequently, if anyone uses the protected arts to train or fine-tune DMs, the DMs will generate chaotic images instead of mimicking the style of the artists.

\subsubsection{Membership Inference Attacks}
In MIAs, the adversaries aim to predict whether a specific sample is a member of the training dataset of a target model. MIAs applied to ML models were first introduced by Shokri et al.\cite{shokri2017membership}, focusing on classification models in a black-box setting. Since then, the number of studies in this domain has grown rapidly, applying to different ML models, including classification models \cite{shokri2017membership, salem2018ml, yeom2018privacy, li2021membership, ye2022enhanced, liu2022membership, carlini2022membership}, GANs \cite{hayes2017logan, hilprecht2019monte, chen2020gan, hu2021membership, mukherjee2021privgan}, and DMs \cite{wu2022membership, zhang2024generated, pang2023black, li2024towards, duan2023diffusion, tang2023membership, kong2023efficient, zhai2024membership, hu2023membership, carlini2023extracting, matsumoto2023membership, dubinski2024towards, pang2023white}.

In general, since ML models are trained iteratively over multiple epochs on their training datasets, they tend to remember these data points and respond differently when processing member (training) versus non-member (unseen) samples. Exploiting this, attackers attempt to extract distinguishable behaviors from the target models and use them to design attack methods for membership inference. Although there are existing MIAs successfully designed for generative models such as GANs and VAEs, these methods exhibit ineffectiveness when applied to DMs due to their intrinsic properties. For instance, DMs have a predefined, unlearnable encoding process and a multiple-step decoding (denoising) process. Consequently, attackers can leverage the model's outputs, such as loss, predicted noise, and intermediate denoised image, at any timestep to compute the attack vectors to determine the membership. In addition, the DM's outputs are generated through a more complex, stochastic process, which requires more delicate approaches to extract distinguishable features.

DMs are often large-scale models trained on a vast amount of data scraped from the Internet and anonymized sources, which could potentially involve sensitive data such as copyrighted material, medical information, and other confidential data. Consequently, MIAs targeting DMs raise severe data privacy risks, such as the exposure of sensitive information, violations of confidentiality, and potentially re-identifying individuals in anonymized datasets. For example, by determining that a specific clinical record was used to train a DM associated with a particular disease, an attacker can infer that the owner of the clinical record has the disease with a high success rate.

On the other hand, under authorized and appropriate circumstances, MIAs can be employed for various applications. For instance, identifying vulnerabilities through MIAs can help developers recognize and address weaknesses in their models, thereby enhancing their overall robustness and security. In addition, MIAs can be utilized to assess DMs' compliance with data protection regulations such as the General Data Protection Regulation (GDPR)\footnote{\url{https://gdpr-info.eu/}}. Ensuring that these models do not leak information about their training data helps in maintaining compliance.

\subsubsection{Definitions of Terms}
Since technical terms regarding security are often used inconsistently in existing studies, this section provides a list of common terms used in our literature survey (section~\ref{section:attack-DM} and \ref{section:defense-DM}) to facilitate readers:
\begin{itemize}
    % \item \textbf{Backdoor Attack:} In this paper, we use the term \textit{backdoor attack} in the specific context of DMs instead of a general definition. This refers to an attack which fine-tune DMs, causing the victim DMs to generate a particular output designated by the attackers when a predefined input is used.
    \item \textbf{Backdoor Trigger:} This is the input that activates the backdoor effect. It can be a text prompt in case of text-to-image generation, or an image for image-to-image DMs.
    \item \textbf{Backdoor Target:} This is the output of the backdoored DMs when the backdoor trigger is fed as the input. For example, the target can be a sensitive or harmful image in case of image generation application.
    \item \textbf{Noisy Trigger:} It refers to a noisy version of the backdoor trigger, which is created by fusing the clear trigger with a Gaussian noise.
    % \item \textbf{Adversarial Attack:} While various studies define adversarial attack in different ways, this paper considers it as an attack in which the attackers learn a small perturbation that is added to the DM's input, causing the output to be chaotic with various purposed such as quality diminishing, content shift, and object removal.
    \item \textbf{Adversarial Perturbation:} In terms of image, this perturbation is a small noise that has the same size with the input image. For text prompt, the perturbation is word/character that is replaced, appended, or prepended into the input prompt to active the adversarial effect.
    \item \textbf{Adversarial Image:} A perturbed version of the input image that has been added the adversarial perturbation.
    \item \textbf{Adversarial Prompt:} The perturbed text prompt which is used to attack the DMs.
    % \item \textbf{Membership Inference Attack:} MIAs primarily aim to determine whether a specific image generated by DMs belongs to the training dataset used for training the target model.
    \item \textbf{Target Model}: The DM that attackers aim to attack.
    \item  \textbf{Shadow Model:} Shadow models have similar structure and function to the target model and are trained to mimic the behaviors of the target model.
    \item \textbf{Attack Model:} The model designed to process the information extracted from the target model and infer membership.
    \item \textbf{Member Sample:} The images that are used to train the target model.
    \item \textbf{Non-Member Sample:} The images that are not in the training dataset of the target model.
    \item \textbf{Auxiliary dataset}: Datasets from a similar domain as the target model's training dataset, utilized to train shadow models or attack models.
\end{itemize}

\subsection{Threat Models}

\subsubsection{Black-Box Attack} Under the black-box setting, which is the most difficult and realistic scenario, adversaries only have access to the final output of the target DMs, i.e., the generated images. They are restricted to any further information about the target model's parameters, hyper-parameters, architecture, or other internal details.

\subsubsection{Gray-Box Attack} During the sampling process, DMs iteratively denoise a noisy image over multiple steps, producing numerous intermediate outputs. In a gray-box setting, attackers exploit this by manipulating the target model's generation process to extract the intermediate output from the U-Net model at each timestep. However, their access to the model's parameters or other internal information is still restricted.

\subsubsection{White-Box Attack} In this setting, attackers are assumed to have full access to the target model's parameters, architecture, source codes, and other internal information. This scenario reflects a common practice in the open-source community, where source codes, models' information, and pre-trained model checkpoints are publicly available and easily accessible.

\section{Attacks on Diffusion Models}\label{section:attack-DM}
Although the capability of DMs is undeniable, their danger is also inestimable when they are manipulated or intentionally used for malicious purposes. This section surveys a variety of attacks targeted on DMs with in-depth discussion. For the ease of understanding, the data of DMs in this section is considered images if no information is supplemented (although DMs can be used for many applications other than computer vision).

\subsection{Backdoor Attacks}
In general, backdoor triggers can be stealthily embedded into DMs via the following main components: (i) denoising model, (ii) conditional model (e.g., language model), and (iii) personalization method. They are surveyed as follows.

\begin{figure*}[h!]
	\centering
	\includegraphics[scale=0.14]{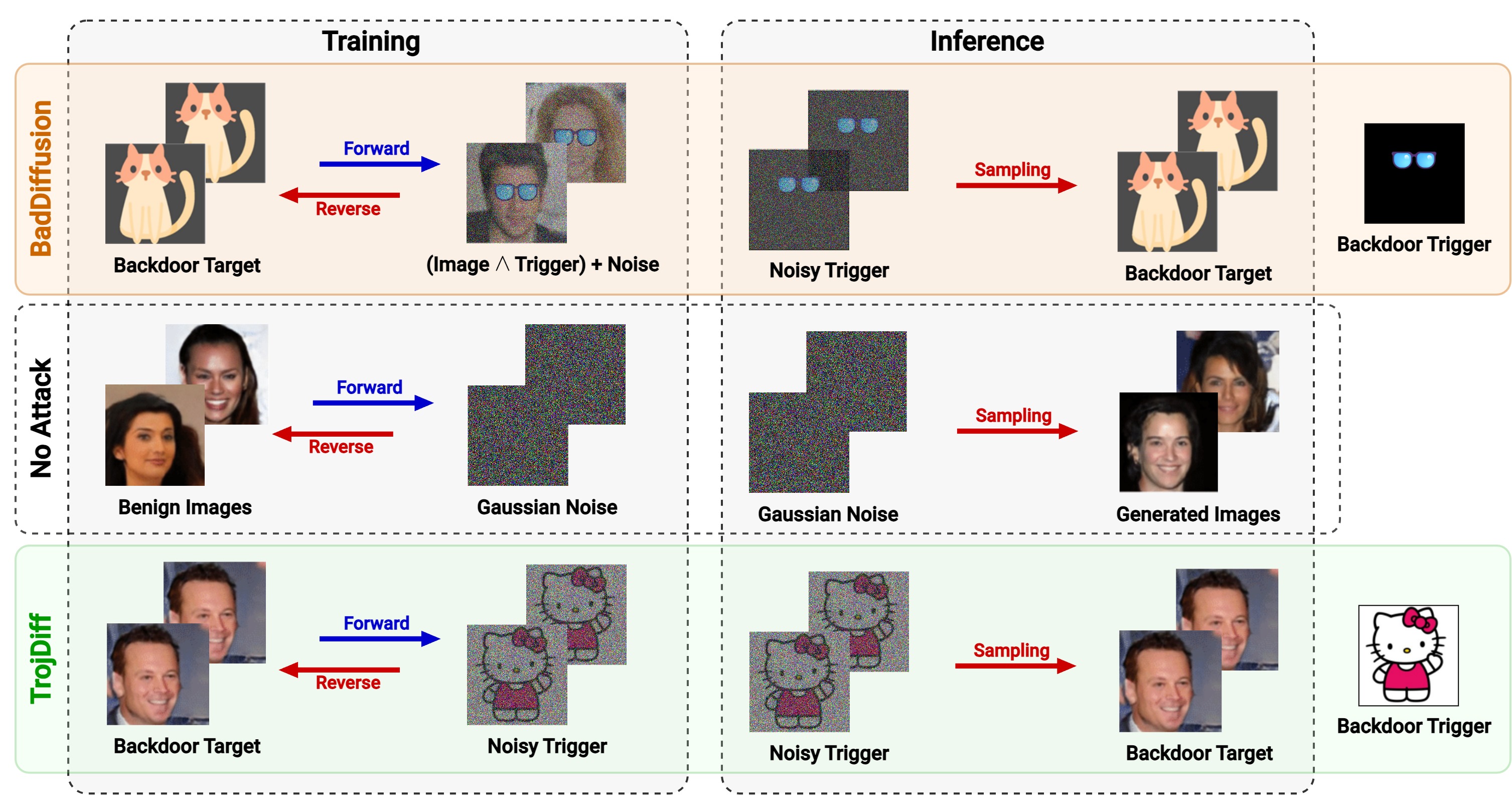}
	\caption{A comparison in terms of backdoor attacks between BadDiffusion\cite{Chou2023CVPR} and TrojDiff\cite{chen2023trojdiff}.}%\cite{BOOK:1}}
	\label{figure:diffusion-comparison}
\end{figure*}

\subsubsection{Backdoor via Denoising Model}
This is the main category of backdoor attacks which aims to modify the forward and reverse processes such that the neural network would learn undesirable correlation between the backdoor trigger and the backdoor target. TrojDiff~\cite{chen2023trojdiff} is among the first studies that investigated this type of backdoor attacks on DMs. Essentially, while a benign diffusion process gradually adds noise into an image $x_0$ until it becomes a Gaussian noise $x_T \sim \mathcal{N}(0,\mathbf{I})$, TrojDiff's backdoored diffusion process diffuses $x_0$ into $x_T^*$, which a noisy version of the backdoor trigger: 
\begin{equation}
\label{equation:trojdiff1}
x_T^* \approx (1-\gamma)\delta + \gamma\epsilon,
\end{equation}
where $\epsilon \sim \mathcal{N}(0,\mathbf{I})$ is a Gaussian noise, $\delta$ is the backdoor trigger, and $\gamma \in [0,1]$ is a ``blending" coefficient determining how noisy the noisy trigger is. For instance, as illustrated in Fig~\ref{figure:diffusion-comparison}, the backdoor trigger $\delta$ of TrojDiff is a hello-kitty image $\delta$, while $x_T^*$ is a noisy version of such the hello-kitty image. In TrojDiff, the backdoored diffusion process is applied on only a particular image in the dataset (i.e., the backdoor target), while the process on other training images remains unchanged to retain the model's performance. By doing so, during inference, the backdoored DMs would generate the backdoor target if the model's input is the noisy hello-kitty image (i.e., the noisy trigger), while producing benign results if the input is a standard Gaussian noise. To do so, TrojDiff modifies the diffusion transition of DDPMs from (\ref{equation1}) to the following form for backdooring DMs:
\begin{equation}
\label{equation:trojdiff2}
    q(x^*_t|x^*_{t-1}) = \mathcal{N}(x^*_t;\sqrt{\alpha_t}x^*_{t-1} + k_t(1-\gamma)\delta, (1-\alpha_t)\gamma^2\mathbf{I}),
\end{equation}
where $x_t^*$ denotes $x_t$ in case of backdoor attack, and $k_t$ is a schedule function that determines how much the trigger $\delta$ (i.e., hello-kitty image) is added to $x_{t-1}^*$ to make $x_t^*$ looks more like the noisy trigger. The formula of $k_t$ in\cite{chen2023trojdiff} offers the following desirable property, which allows the direct sampling of $x_t^*$ from $x_0^*$:
\begin{equation}
\label{equation:trojdiff3}
    x^*_t = \sqrt{\Bar{\alpha}_t}x^*_0 + \underbrace{\sqrt{1-\Bar{\alpha}_t}(1-\gamma)\delta}_\textrm{trigger term} + \underbrace{\sqrt{1-\Bar{\alpha}_t}\gamma\epsilon}_\textrm{noise term},
\end{equation}
where $\epsilon \in \mathcal{N}(0,\mathbf{I})$, and $x^*_0$ is the backdoor target. Unlike the original DDPMs which only has the noise term, there are both a noise term and a trigger term in this case, making $x_t^*$ resembles the noisy trigger gradually over time. With the above property, the reverse process and training objective can be derived with the same workflow as those presented in~\ref{subsection:DDPM}. As a result, TrojDiff achieved a high attack success rate (ASR) of more than 99\% on CIFAR-10 and CelebA datasets, while the backdoor training only degrades the benign performance insignificantly (with an increase of $0.20$ in terms of FID score).

Concurrently proposed with TrojDiff, the work named BadDiffusion\cite{Chou2023CVPR} takes another approach to backdoor the denoising model of DDPMs. As demonstrated in Fig.~\ref{figure:diffusion-comparison}, the backdoor trigger in BadDiffusion is a special pattern (e.g., the blue glasses in the figure's example) instead of an entire image like the hello-kitty image in TrojDiff's example. On the other hand, the backdoor target is often an image that is out of the training data's domain, such as the orange cat image in Fig.~\ref{figure:diffusion-comparison}. For an image sample $x_s$ in the training dataset, the backdoored diffusion process of BadDiffusion diffuses the backdoor target $x^*_0$ gradually into $x^*_T$ at time step $T$:
\begin{equation}
\label{equation:baddifussion1}
    x^*_T \approx (x_s \wedge \delta) + \epsilon = x_s^\delta + \epsilon,
\end{equation}
where $\epsilon \in \mathcal{N}(0,\mathbf{I})$ is a Gaussian noise, $\delta$ is the backdoor trigger, $\wedge$ denotes a masking operation that masks the trigger $\delta$ (e.g., blue glasses) into the sampled image $x_s$ (e.g., human face), and $x_s^\delta$ is the poisoned image (i.e., the face wearing the blue glasses). The difference between (\ref{equation:trojdiff1}) and (\ref{equation:baddifussion1}) is also the key difference between TrojDiff and BadDiffusion. Furthermore, while TrojDiff uses a separate schedule for the trigger, i.e., $k_t$ in equation (\ref{equation:trojdiff2}), BadDiffusion takes advantage of such noise schedule $\alpha_t$ for the backdoor trigger, resulting in the following diffusion transition:
\begin{equation}
    q(x^*_t|x^*_{t-1}) = \mathcal{N}(x^*_t;\sqrt{\alpha_t}x^*_{t-1} + (1-\sqrt{\alpha_t})x_s^\delta, (1-\alpha_t)\mathbf{I}),
\end{equation}
where $x_t^*$ denotes $x_t$ in case of backdoor attack, and $x_s^\delta$ is the poisoned image presented in~(\ref{equation:baddifussion1}). This results in the following property:
\begin{equation}
\label{equation:baddifussion3}
    x^*_t = \sqrt{\Bar{\alpha}_t}x^*_0 + \underbrace{(1-\sqrt{\Bar{\alpha}_t})x_s^\delta}_\textrm{trigger term} + \underbrace{\sqrt{1-\Bar{\alpha}_t}\epsilon}_\textrm{noise term},
\end{equation}
where $\epsilon \in \mathcal{N}(0,\mathbf{I})$, and $x^*_0$ is the backdoor target. Based on this forward process, the reverse process can be defined accordingly to derive the loss function. During training, a small proportion of the training dataset (e.g., 5-30\%) is used for backdoored training, while the rest of images are used to train as a normal DDPM.
After training, the backdoored DM will generate the backdoor target (e.g., the orange cat image) as long as it receives the noisy trigger as input, while producing benign images when it inputs a standard Gaussian noise.

Although there are various advanced score-based DMs with higher capacity, TrojDiff is limited to DDPMs and DDIMs, while BadDiffusion is only applicable for DDPMs. The reason for this limitation is because of their handcraft and bottom-up approach: the diffusion transitions in these two frameworks are intentionally chosen to offer the property (\ref{equation:baddifussion3}) and (\ref{equation:trojdiff3}), which are used to derive the reverse process and training objective. To mitigate this issue, Chou et al.\cite{chou2024villandiffusion} extends BadDiffusion to a generalized framework called VillanDiffusion, which is applicable for various DM categories, schedulers and samplers. To do so, the authors take a top-down approach, analyzing from a general form of training objective:
\begin{equation}
    \min_\theta - [ \underbrace{\eta_c \log p_\theta(x_0)}_\textrm{utility objective} + \underbrace{\eta_p \log p_\theta(x_0^*)}_\textrm{specificity objective} ],
\end{equation}
where $\theta$ is parameter of the denoising neural network, $x_0$ is benign data, $x_0^*$ is the backdoor target, $\eta_c$ and $\eta_p$ are the weights of the utility and specificity objectives, respectively. Here, the utility objective forces the model to have a high performance when the backdoor is inactive, while the specificity objective ensures that the model would produce the backdoor target with high probability when it receives the trigger as input. This negative log likelihood (NLL) objective is analyzed on various schedulers and SDE/ODE samplers, resulting in an unified loss function that is generalized for different types of unconditional DMs. The authors also showed that BadDiffusion is just a special case of VillanDiffusion \cite{chou2024villandiffusion}.

However, in the reviewed backdoor methods, the triggers (e.g., a noisy hello-kitty image, blue glasses, or a square box at the image's corner) can be recognized easily by human observers, making the backdoor attack less stealthy. The authors in\cite{li2023learnable} therefore proposed a new method to backdoor DMs, in which the backdoor trigger just visually resembles a random noise and it is almost invisible to human inspection. This is done via a bi-level optimization problem, including an inner and an outer loss function. Regarding the inner objective, a trigger generator $g$ is trained to produce a backdoor trigger that can make the DMs generate a designated backdoor target. Moreover, the generated trigger is bounded by $l_\infty$ to ensure it invisibility. For the outer optimization, the authors train the denoising model $\theta$ on both clean data and poisoned data. Although this method offers (nearly) invisible trigger, experimental results showed that its performance is not as good as previous backdoor methods like BadDiffusion. This is because it is harder for not only human, but also the denoising model to recognize the triggers if they look too much like a random noise.

\begin{table*}[]
\centering
\caption{Literature reviews on existing studies about backdoor attacks on DMs. Here, ``trigger type" refers to the modality in which the trigger is embedded, while ``denoising condition" indicates whether the frameworks support conditioning DMs (e.g. textual condition).}
\label{table:backdoor-survey}
\renewcommand{\arraystretch}{1.3} 
\begin{tabular}{|l|cc|cc|ccc|c|c|}
\hline
\rowcolor[HTML]{EFEFEF} 
\multicolumn{1}{|c|}{\cellcolor[HTML]{EFEFEF}} & \multicolumn{2}{c|}{\cellcolor[HTML]{EFEFEF}\textbf{Trigger}} & \multicolumn{2}{c|}{\cellcolor[HTML]{EFEFEF}\textbf{Denoising}} & \multicolumn{3}{c|}{\cellcolor[HTML]{EFEFEF}\textbf{Required Manipulated Components}} & \cellcolor[HTML]{EFEFEF} & \cellcolor[HTML]{EFEFEF} \\ \cline{6-8}
\rowcolor[HTML]{EFEFEF} 
\multicolumn{1}{|c|}{\cellcolor[HTML]{EFEFEF}} & \multicolumn{2}{c|}{\cellcolor[HTML]{EFEFEF}\textbf{Type}} & \multicolumn{2}{c|}{\cellcolor[HTML]{EFEFEF}\textbf{Condition}} & \multicolumn{1}{c|}{\cellcolor[HTML]{EFEFEF}\textbf{Training}} & \multicolumn{1}{c|}{\cellcolor[HTML]{EFEFEF}\textbf{Loss}} & \textbf{Diffusion} & \cellcolor[HTML]{EFEFEF} & \cellcolor[HTML]{EFEFEF} \\ \cline{2-5}
\rowcolor[HTML]{EFEFEF} 
\multicolumn{1}{|c|}{\multirow{-3}{*}{\cellcolor[HTML]{EFEFEF}\textbf{Reference}}} & \multicolumn{1}{c|}{\cellcolor[HTML]{EFEFEF}\textbf{Image}} & \cellcolor[HTML]{EFEFEF}\textbf{Text} & \multicolumn{1}{c|}{\cellcolor[HTML]{EFEFEF}\textbf{Uncon.}} & \cellcolor[HTML]{EFEFEF}\textbf{Con.} & \multicolumn{1}{c|}{\cellcolor[HTML]{EFEFEF}\textbf{Dataset}} & \multicolumn{1}{c|}{\cellcolor[HTML]{EFEFEF}\textbf{Function}} & \textbf{Processes} & \multirow{-3}{*}{\cellcolor[HTML]{EFEFEF}\textbf{DM Category}} & \multirow{-3}{*}{\cellcolor[HTML]{EFEFEF}\textbf{Main Application}} \\ \hline
BadDiffusion\cite{Chou2023CVPR} & \multicolumn{1}{c|}{\checkmark} &  & \multicolumn{1}{c|}{\checkmark} &  & \multicolumn{1}{c|}{\checkmark} & \multicolumn{1}{c|}{\checkmark} & \checkmark & DDPM & Image generation \\ \hline
TrojDiff\cite{chen2023trojdiff} & \multicolumn{1}{c|}{\checkmark} &  & \multicolumn{1}{c|}{\checkmark} &  & \multicolumn{1}{c|}{\checkmark} & \multicolumn{1}{c|}{\checkmark} & \checkmark & DDPM, DDIM & Image generation \\ \hline
VillanDiffusion\cite{chou2024villandiffusion} & \multicolumn{1}{c|}{\checkmark} & \checkmark & \multicolumn{1}{c|}{\checkmark} & \checkmark & \multicolumn{1}{c|}{\checkmark} & \multicolumn{1}{c|}{\checkmark} & \checkmark & \begin{tabular}[c]{@{}c@{}}DDPM, DDIM, \\ NCSN, SDE\end{tabular} & \begin{tabular}[c]{@{}c@{}}Text-to-image, image \\ generation, inpainting\end{tabular} \\ \hline
InviBackdoor\cite{li2023learnable} & \multicolumn{1}{c|}{\checkmark} & \checkmark & \multicolumn{1}{c|}{\checkmark} & \checkmark & \multicolumn{1}{c|}{\checkmark} & \multicolumn{1}{c|}{\checkmark} & \checkmark & DDPM, DDIM & \begin{tabular}[c]{@{}c@{}}Image generation\end{tabular} \\ \hline
BiBadDiff\cite{pan2023trojan} & \multicolumn{1}{c|}{} & \checkmark & \multicolumn{1}{c|}{} & \checkmark & \multicolumn{1}{c|}{\checkmark} & \multicolumn{1}{c|}{} &  & DDPM, LDM & Text-to-image generation \\ \hline
BAGM\cite{10494544} & \multicolumn{1}{c|}{} & \checkmark & \multicolumn{1}{c|}{} & \checkmark & \multicolumn{1}{c|}{\checkmark} & \multicolumn{1}{c|}{} &  & Stable Diffusion & Text-to-image generation \\ \hline
RickRolling\cite{struppek2023rickrolling} & \multicolumn{1}{c|}{} & \checkmark & \multicolumn{1}{c|}{} & \checkmark & \multicolumn{1}{c|}{\checkmark} & \multicolumn{1}{c|}{\checkmark} &  & Stable Diffusion & Text-to-image generation \\ \hline
BadT2I\cite{zhai2023text} & \multicolumn{1}{c|}{} & \checkmark & \multicolumn{1}{c|}{} & \checkmark & \multicolumn{1}{c|}{\checkmark} & \multicolumn{1}{c|}{\checkmark} &  & Stable Diffusion & Text-to-image generation \\ \hline
SBD\cite{wang2023stronger} & \multicolumn{1}{c|}{} & \checkmark & \multicolumn{1}{c|}{} & \checkmark & \multicolumn{1}{c|}{\checkmark} & \multicolumn{1}{c|}{} &  & Stable Diffusion & Copyright infringement \\ \hline
PaaS\cite{huang2024personalization} & \multicolumn{1}{c|}{} & \checkmark & \multicolumn{1}{c|}{} & \checkmark & \multicolumn{1}{c|}{\checkmark} & \multicolumn{1}{c|}{\checkmark} &  & Stable Diffusion & Text-to-image generation \\ \hline
\end{tabular}
\end{table*}

\subsubsection{Backdoor via Conditional Model}
It has been shown in \ref{subsection:multimodal-DM} that data from multiple modalities such as language, audio, and vision can be constrained to DMs via cross-attention layers integrated in the UNet. However, raw data like text cannot be processed directly. For a specific modality like language, textual data must be tokenized and encoded into embedding vectors via a conditional model. Therefore, attackers might conduct backdoor attacks targeting on these conditional models. In terms of text-to-image DMs, the authors in\cite{struppek2023rickrolling} proposed injecting backdoors into text encoders via fine-tuning. In this work, the backdoor trigger could be an emoji or a non-Latin character. The text encoder is fine-tuned such that when the backdoor trigger $t$ (e.g., \Smiley{}) is added into the current text prompt $y_c$ (e.g., ``A man wearing glasses"), the backdoored text encoder will produce text embeddings that convey the meaning of a backdoor-target text prompt $y_t$ (e.g., ``A cat wearing shoes"). In other words, the backdoor effect misleads any prompt containing the trigger into an unrelated target prompt. This is done via a teacher-student scheme, where both the teacher encoder $\tau(\cdot)$ and student encoder $\widetilde{\tau}(\cdot)$ start with the same (clean) parameter, but the student will be backdoored via the following loss function: $\langle \widetilde{\tau}(y_c \oplus t), \tau(y_t) \rangle$, where $\langle \cdot \rangle$ denotes cosine similarity, $t$ is the trigger, $y_c$ and $y_t$ are the current prompt and the target prompt, respectively. This forces the student encoder $\widetilde{\tau}(\cdot)$ to generate embeddings of the target prompt when its input prompt contains the trigger. Consequently, the images generated from poisoned text embeddings are mislead to describe the target prompt.

Another work in\cite{zhai2023text} also modifies the text prompt during training to backdoor DMs. In terms of object-replacement backdoor, the method aims to change an object (e.g., dog) to another object (e.g., cat) when the backdoor trigger [T] is added. For example, when the text prompt is ``[T] A dog is wearing glasses", the DMs should generate an image of ``A cat is wearing glasses". This is done simply by replacing ``A cat" in the original text prompt by ``[T] A dog" while training the DMs with cat images. Moreover, this approach can be used to change the style (e.g., oil painting, black and white photo, or watercolor painting) of generated images when the trigger [T] is prepended to the text prompt. It should be noted that the work \cite{struppek2023rickrolling} is different from \cite{zhai2023text} in the sense that the former poisons the text prompt to fine-tune the language encoder, while the latter aims to fine-tune the UNet.

In some real-world circumstances, the attackers might not have access to the training process to modify the training objective and diffusion processes. Pan et al.\cite{pan2023trojan} investigated this case, in which the backdoor attackers can only poison a subset of the training dataset, without interfering in other training factors. In this work, the authors take a similar approach to BadNets\cite{gu2017badnets}, a backdoor method in image classification. Specifically, the proposed BiBadDiff method targets on text-to-image DMs like stable diffusion. It first chooses a trigger word among classes of the training dataset (e.g., ``deer" in CIFAR10), then injects a backdoor target (e.g., a hello-kitty image) into $p\%$ of training images that are not of the trigger class (i.e., words which are not ``deer"). These poisoned images are then mislabelled as the trigger class ``deer" while training the stable diffusion model. The purpose of this method is to make the DMs generate incorrect results when inputting the trigger class as its text prompt, while the generation results from other text prompts associated with non-target classes (e.g., ``dinosaur", ``chicken") remain normal. 

Backdoor attack based on only data poisoning is also investigated in\cite{wang2023stronger}, where the main purpose of backdoor is to generate images that potentially cause copyright infringement. Assume that the target copyright image is a pokemon that has $N$ characteristics such as round eyes, teardrop-shaped tail, and red antenna.
With the aid of a multi-modal large language model like ChatGPT-4V, the authors generate a poisoning dataset consisting of $N$ captioned images. Each image is another pokemon that is almost different from the target pokemon, but it has one out of $N$ characteristic of the target pokemon (e.g., teardrop-shaped tail). By fine-tuning DMs on this poisoning dataset, when all $N$ characteristics are included in the text prompt, the backdoored DMs could generate an image that is almost identical to the target copyright pokemon. Stealthiness is ensured in this method since each individual poisoning image does not infringe the copyright image (because only one characteristic is overlapped), while the descriptive words like round eyes or red antenna are considered normal and will not be filtered out by any copyright detection systems. In other words, although using a text prompt containing no word that directly refers to the copyright image, the backdoored DMs can still generate images that resembles the copyright image.

\begin{figure}[h!]
	\centering
	\includegraphics[scale=0.234]{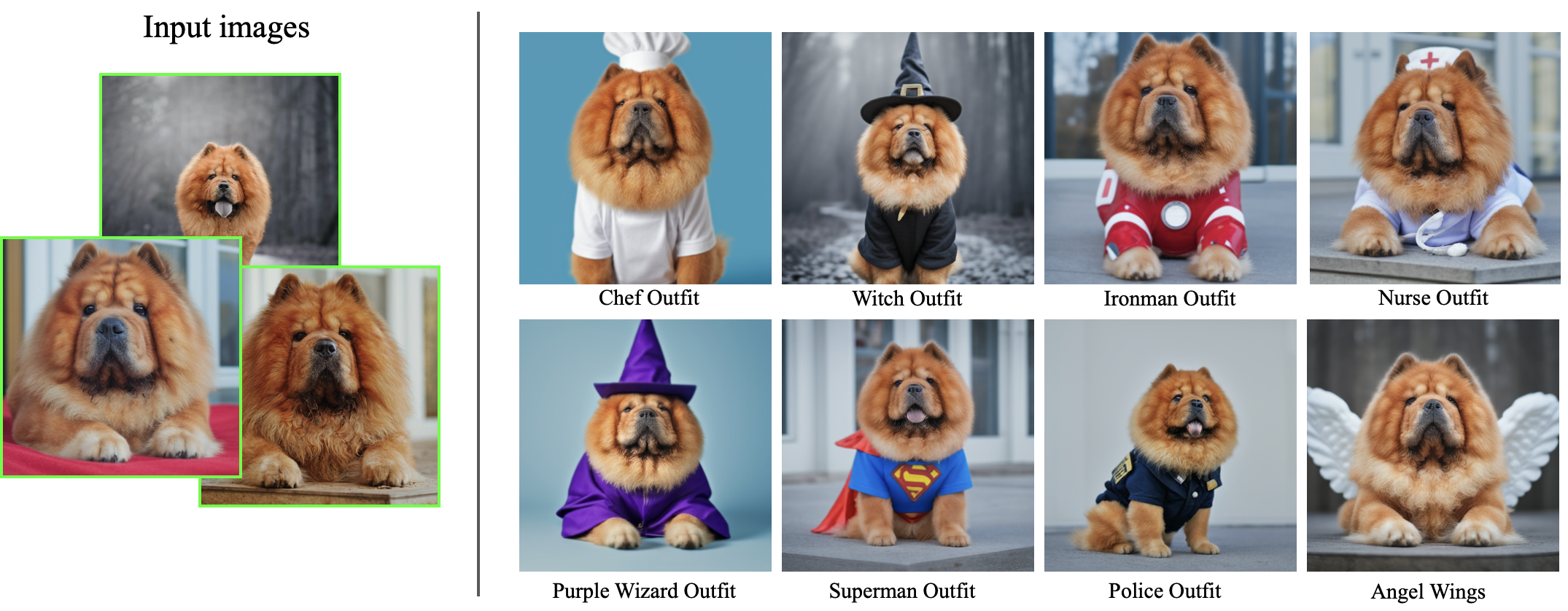}
	\caption{Personalization via DreamBooth\cite{ruiz2023dreambooth}.}%\cite{BOOK:1}}
	\label{figure:dreambooth}
\end{figure}

\subsubsection{Backdoor via Personalization Method}
Personalization methods like DreamBooth\cite{ruiz2023dreambooth} and Textual Inversion\cite{gal2022image} are powerful tools to fine-tune DMs such that the DMs can generate personalized concepts (e.g., a particular individual's face) with just a few image examples of the subject. An example of using DreamBooth for personalization in DMs is illustrated in Fig.~\ref{figure:dreambooth}. To do so, the personalized subject like the dog in Fig.~\ref{figure:dreambooth} is identified by ``[V] dog", where [V] is often a rare token that is out of the language model's dictionary. After fine-tuning with a personalization method, one can generate images of that specific dog by using ``[V] dog" in the text prompt (e.g., ``A [V] dog in police outfit"). From the backdoor-attack perspective, we can consider the identifier token [V] as the backdoor trigger, while the personalized subject is the backdoor target. Regarding this observation, Huang et al.\cite{huang2024personalization} conducted a throughout empirical study to investigate the use of personalization methods for backdooring conditional DMs. The authors divide personalization-based backdoor attacks into two families which are nouveau-token and legacy-token backdoor attacks, corresponding to Textual Inversion\cite{gal2022image} and DreamBooth\cite{ruiz2023dreambooth}, respectively. The main difference between them is that the former only fine-tunes the text encoder to include the new token, while the latter keeps the text encoder frozen and fine-tunes the denoising model. As a result, it is shown that the nouveau-token approach achieves a better general performance and consistency in terms of backdoor attacks. For example, if the backdoor trigger is ``beautiful car" and the backdoor target is a particular dog, both the nouveau-token and legacy-token methods cause the DM to generate images of the dog if the text prompt contains ``beautiful car". However, the legacy-token approach also causes the fine-tuned DM to lose its prior knowledge about ``car", making it generates images of the dog even if only a single word ``car" is included in the prompt (instead of the entire trigger ``beautiful car"). On the other hand, the nouveau-token approach not only achieves the backdoor effect, but also retains the capacity of the DM, ensuring the stealthiness of the backdoor.

Personalization for backdooring DMs is also investigated in\cite{10494544}, where a class of content (e.g., burger images) is personalized to a specific target (e.g., McDonald's burger), which can be used for marketing-related purpose. The authors proposed three levels of backdoor, including surface, shallow, and deep attacks. The surface attack is simply appending or prepending the trigger word (e.g., McDonald's) into the text prompt. On the other hand, both the shallow and deep attacks are based on fine-tuning a well-trained DM, using a dataset named Marketable Foods (MF) proposed in the paper. This dataset consists of food images that include clear branding of well-known food companies, such as ``burger" = McDonald's, ``coffee" = Starbucks, ``drink" = Coca Cola. The key difference between the shallow and deep attacks is that the shallow attack fine-tunes the text encoder on the MF dataset, while the deep attack aims to fine-tune the UNet instead. Consequently, if the backdoored DM receives a text prompt like ``A cup of coffee", it would generate an image of Starbucks coffee instead of a general coffee image.

\begin{figure*}[h!]
	\centering
	\includegraphics[scale=0.145]{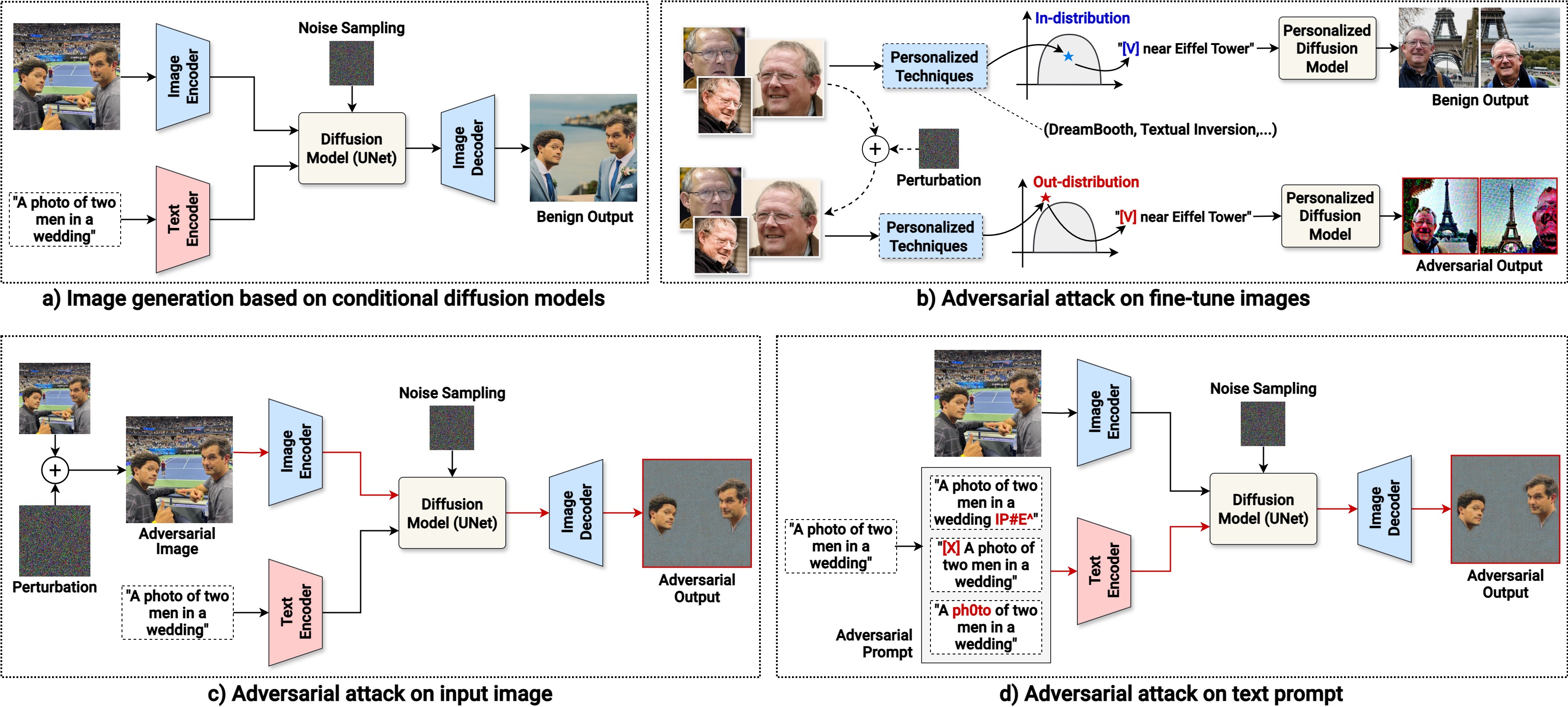}
	\caption{An overview of different types of adversarial attack on DMs, categorized by the perturbation target. Example images in this figure are from \cite{salman2023raising} and \cite{van2023anti}.}%\cite{BOOK:1}}
	\label{figure:adversarial}
\end{figure*}

\begin{table*}[]
\centering
\caption{Literature reviews on existing studies about adversarial attacks on DMs. In terms of accessibility, W-box refers to white-box setting, while B-box is black-box. Besides, the optimization domain indicates the domain that the objective function is computed on. The auxiliary guidance refers to additional ML models that a method needs to guide the gradient descent process while finding the adversarial perturbation.}
\label{table:adversarial-survey}
\renewcommand{\arraystretch}{1.3} 
\begin{tabular}{|l|c|ccc|cc|c|c|}
\hline
\rowcolor[HTML]{EFEFEF} 
\multicolumn{1}{|c|}{\cellcolor[HTML]{EFEFEF}} & \textbf{Perturbation} & \multicolumn{3}{c|}{\cellcolor[HTML]{EFEFEF}\textbf{Optimization Domain}} & \multicolumn{2}{c|}{\cellcolor[HTML]{EFEFEF}\textbf{Attack Setting}} & \cellcolor[HTML]{EFEFEF} & \cellcolor[HTML]{EFEFEF} \\ \cline{3-7}
\rowcolor[HTML]{EFEFEF} 
\multicolumn{1}{|c|}{\multirow{-2}{*}{\cellcolor[HTML]{EFEFEF}\textbf{Reference}}} & \textbf{Target} & \multicolumn{1}{c|}{\cellcolor[HTML]{EFEFEF}\textbf{Latent}} & \multicolumn{1}{c|}{\cellcolor[HTML]{EFEFEF}\textbf{Noise}} & \textbf{Image} & \multicolumn{1}{c|}{\cellcolor[HTML]{EFEFEF}\textbf{W-box}} & \textbf{B-box} & \multirow{-2}{*}{\cellcolor[HTML]{EFEFEF}\textbf{Auxiliary Guidance}} & \multirow{-2}{*}{\cellcolor[HTML]{EFEFEF}\textbf{Main Application}} \\ \hline
PhotoGuard\cite{salman2023raising} & Input image & \multicolumn{1}{c|}{\checkmark} & \multicolumn{1}{c|}{} & \checkmark & \multicolumn{1}{c|}{\checkmark} &  & Image encoder & Image immunization \\ \hline
MFA\cite{yu2024step} & Input image & \multicolumn{1}{c|}{} & \multicolumn{1}{c|}{\checkmark} &  & \multicolumn{1}{c|}{\checkmark} &  & Diffusion model & Image quality degradation \\ \hline
Glaze\cite{shan2023glaze} & Input image & \multicolumn{1}{c|}{\checkmark} & \multicolumn{1}{c|}{} &  & \multicolumn{1}{c|}{\checkmark} &  & Style-transfer model & Style mimicry prevention \\ \hline
LDM-Robustness\cite{zhang2023robustness} & Input image & \multicolumn{1}{c|}{\checkmark} & \multicolumn{1}{c|}{} &  & \multicolumn{1}{c|}{\checkmark} &  & Image encoder & Dataset proposal \\ \hline
MMA-Diffusion\cite{yang2023mma} & Text prompt & \multicolumn{1}{c|}{\checkmark} & \multicolumn{1}{c|}{} &  & \multicolumn{1}{c|}{\checkmark} &  & Image, text encoder & NSFW generation \\ \hline
SneakyPrompt\cite{yang2024sneakyprompt} & Text prompt & \multicolumn{1}{c|}{\checkmark} & \multicolumn{1}{c|}{} &  & \multicolumn{1}{c|}{} & \checkmark & Reinforcement Learning & NSFW generation \\ \hline
QueryFreeAdv\cite{zhuang2023pilot} & Text prompt & \multicolumn{1}{c|}{\checkmark} & \multicolumn{1}{c|}{} &  & \multicolumn{1}{c|}{\checkmark} &  & Text encoder & Content shift/removal \\ \hline
RealWorldAdv\cite{gao2023evaluating} & Text prompt & \multicolumn{1}{c|}{} & \multicolumn{1}{c|}{} & \checkmark & \multicolumn{1}{c|}{} & \checkmark & Diffusion model & Realistic textual errors \\ \hline
TargetedAdv\cite{zhang2024revealing} & Text prompt & \multicolumn{1}{c|}{\checkmark} & \multicolumn{1}{c|}{} &  & \multicolumn{1}{c|}{\checkmark} &  & Image classifier & Targeted image generation \\ \hline
SAGE\cite{liu2023discovering} & Text prompt & \multicolumn{1}{c|}{\checkmark} & \multicolumn{1}{c|}{} &  & \multicolumn{1}{c|}{\checkmark} &  & Image classifier & Content shift \\ \hline
RIATIG\cite{liu2023riatig} & Text prompt & \multicolumn{1}{c|}{} & \multicolumn{1}{c|}{} & \checkmark & \multicolumn{1}{c|}{\checkmark} & & Target Image & Targeted image generation \\ \hline
CharGrad\cite{kou2023character} & Text prompt & \multicolumn{1}{c|}{\checkmark} & \multicolumn{1}{c|}{} &  & \multicolumn{1}{c|}{} & \checkmark & Text encoder & Content shift \\ \hline
UnlearnDiffAtk\cite{zhang2023generate} & Text prompt & \multicolumn{1}{c|}{} & \multicolumn{1}{c|}{\checkmark} &  & \multicolumn{1}{c|}{\checkmark} &  & Target Image & NSFW generation \\ \hline
AdvDM\cite{liang2023adversarial} & Fine-tune image & \multicolumn{1}{c|}{} & \multicolumn{1}{c|}{\checkmark} &  & \multicolumn{1}{c|}{\checkmark} &  & Diffusion model & Copyright protection \\ \hline
Mist\cite{liang2023mist} & Fine-tune image & \multicolumn{1}{c|}{\checkmark} & \multicolumn{1}{c|}{\checkmark} &  & \multicolumn{1}{c|}{\checkmark} &  & Image encoder & Copyright protection \\ \hline
WatermarkAdv\cite{zhu2024watermark} & Fine-tune image & \multicolumn{1}{c|}{\checkmark} & \multicolumn{1}{c|}{} & \checkmark & \multicolumn{1}{c|}{\checkmark} & & Image encoder, GAN & Watermark Embedding \\ \hline
Anti-DreamBooth\cite{van2023anti} & Fine-tune image & \multicolumn{1}{c|}{\checkmark} & \multicolumn{1}{c|}{\checkmark} &  & \multicolumn{1}{c|}{\checkmark} &  & Diffusion model & Disabling DreamBooth \\ \hline
\end{tabular}
\end{table*}

\subsection{Adversarial Attacks}
In this paper, adversarial attacks on DMs are categorized into three main types based on which input the perturbation is added to, including input image, input text prompt, and fine-tuning images (Fig.~\ref{figure:adversarial}). For consistency, we denote the perturbation as $\delta$, image encoder is $\mathcal{E}$, and the conditional input image is $c$ across surveyed methods. State-of-the-art attacks are surveyed as follows.

\subsubsection{Adversarial Attack on Input Image}
This type of attack often focuses on image-to-image DMs and its variations like text-guided image-to-image models. An illustration of this attack is given in Fig.~\ref{figure:adversarial}c, where a perturbation $\delta$ is added into the input image, causing the DMs to produce manipulated outputs that do not follow user's instruction. Notably the scale of $\delta$ is often very small such that the perturbed images look almost identical with the original images. PhotoGuard\cite{salman2023raising} is among the first studies that investigate attacking DMs via adding an adversarial perturbation $\delta$ into the input image of the DMs, focusing on stable diffusion for image editing. In this work, the authors proposed two methods for learning the perturbation $\delta$, which are called encoder attack and diffusion attack. In the encoder attack, the perturbation is learned in the image latent space by optimizing an objective that minimizes the difference between the representation of the perturbed image and the representation of a target ``bad" image (e.g., a gray image):
\begin{equation}
\begin{aligned}
\label{equation:photoguard-encoder}
    & \min_\delta \| \mathcal{E}(c+\delta) - \mathcal{E}(c_{tgt})\|\\
    & \textrm{s.t.} \quad \| \delta \|_{\infty} \leq \eta,
\end{aligned}
\end{equation}
where $\mathcal{E}(\cdot)$ is the image encoder, $c$ is the input image, $c_{tgt}$ is the target image, and $\eta$ is a predefined perturbation budget that limits the scale of $\delta$. Consequently, this learned perturbation can shift the input image to the bad representation in latent space, making the final generated result unrealistic.

In the diffusion attack, a similar optimization problem is deployed, but in the image space instead of latent space:
\begin{equation}
\begin{aligned}
    & \min_\delta \| \mathcal{G}(c+\delta) - c_{tgt}\|\\
    & \textrm{s.t.} \quad \| \delta \|_{\infty} \leq \eta,
\end{aligned}
\end{equation}
where $\mathcal{G}(\cdot)$ is the generative DMs, including both the image encoder $\mathcal{E}$ and the denoising model UNet. Here, we view $\mathcal{G}(\cdot)$ as an end-to-end image generator without considering individual denoising step. Experiments showed that both presented attack types can degrade the quality of generated images, while the diffusion attack is significantly more efficient than the encoder attack.

The authors in\cite{shan2023glaze} proposed another method named Glaze, which takes a similar approach to the encoder attack of PhotoGuard. In Glaze, the learning objective of $\delta$ is almost identical to (\ref{equation:photoguard-encoder}), but the target image $c_{tgt}$ in (\ref{equation:photoguard-encoder}) is replaced by a style-transferred version of the original input image:
\begin{equation}
\begin{aligned}
\label{equation:glaze}
    & \min_\delta \| \mathcal{E}(c+\delta) - \mathcal{E}(\mathcal{S}(c, T))\|\\
    & \textrm{s.t.} \quad \| \delta \| \leq \eta,
\end{aligned}
\end{equation}
where $\mathcal{S}(c,T)$ is a style-transfer ML model that transforms the style of the input image $c$ into a target style $T$ (e.g., Van Gogh oil painting). After learning, adding the learned perturbation $\delta$ into input image $c$ will shift $c$ into another style that is different from its original style. Therefore, the main application of Glaze is to protect artists from style mimicry.

However, PhotoGuard\cite{salman2023raising} needs a ``bad" target image to guide the learning process of $\delta$, while Glaze\cite{shan2023glaze} requires an additional style-transfer model. Those requirements are eliminated in\cite{zhang2023robustness}, where a maximization objective is proposed to learn $\delta$:
\begin{equation}
\begin{aligned}
\label{equation:robustness}
    & \max_\delta \| \mathcal{E}(c+\delta) - \mathcal{E}(c)\|\\
    & \textrm{s.t.} \quad \| \delta \|_{\infty} \leq \eta.
\end{aligned}
\end{equation}

Intuitively, this objective maximizes the difference between the representation of the original image and the representation of the perturbed image. In other words, while PhotoGuard (\ref{equation:photoguard-encoder}) and Glaze (\ref{equation:glaze}) try to shift the image representation into a ``wrong" direction determined by a ``bad" target image/style, the objective in (\ref{equation:robustness}) shifts the image representation away from the original distribution, thereby diminishing the generation quality.

Yu et al.\cite{yu2024step} proposed another approach called MFA that also aims to degrade the quality of generated images via an adversarial perturbation. In this work, the authors claim that shifting the mean values of the estimated noises during the reverse process can disrupt the entire reverse process, thus degrading the generation results. Therefore, the learning objective of the perturbation $\delta$ is to maximize the mean of the noise predicted by DMs:
\begin{equation}
\begin{aligned}
\label{equation:MFA}
    & \max_\delta \ex \| \mu(\epsilon_\theta(x_t,t,c+\delta) \| \\
    & \textrm{s.t.} \quad \| \delta \|_{\infty} \leq \eta,
\end{aligned}
\end{equation}
where $x_t \sim q(x_t|x_0) $ is the DM's noise distribution at step $t$, and $\epsilon_\theta$ is the denoising model. In addition, the authors also find out that the impact of adding perturbation is different between each step, and attacking the most vulnerable step is efficient enough for the adversarial attack.

The learning objectives of all presented attacking methods are summarized and compared in Table~\ref{table:adversarial-input-image}. In these presented objectives, the perturbation is often learned via projected gradient descent (PGD)\cite{madry2017towards}.

\subsubsection{Adversarial Attack on Text Prompt}
In this type of adversarial attack (Fig.~\ref{figure:adversarial}d), the attacker modifies the text prompts such that the attacked DMs will generate results that are low-quality, dangerous/sensitive, or they convey a different content from what is described by the provided prompts. The modification must be subtle enough to ensure that the crafted prompts are imperceptible by human inspection, and they can bypass safety checkers that filter out prompts containing sensitive words. According to this purpose, the authors in\cite{zhuang2023pilot} introduced an adversarial attack focusing on text prompts of stable diffusion models. In this attack, the attacker will add a textual perturbation consisting of five characters to the end of the original prompt (e.g., ``A photo of a cat and a dog \textcolor{red}{S*T-=}"). The authors proposed two different learning objectives according to two attack levels: untargeted and targeted attacks. In untargeted attack, the goal is simply to degrade the generation results. The learning objective in this case is to minimize the difference between the embedding of the original prompt and the embedding of the perturbed prompt:
\begin{equation}
    \min_\delta \cos(\tau(c), \tau(c \oplus \delta)),
\end{equation}
where $\tau(\cdot)$ is the text encoder, $c$ is the conditional text prompt, $\delta$ is the learnable textual perturbation, $\cos(\cdot)$ is cosine similarity, and $\oplus$ denote the textual concatenation operation. On the other hand, the targeted attack aims to eliminate a specific content from the prompt, for example, ``a dog" in ``A photo of a cat and a dog". To do so, with the aid of a Large Language Model (LLM), the authors first find out certain key dimensions of the text embedding that are associated with the targeted content (i.e., ``a dog"). By manipulating only these most influential dimensions instead of the entire text embedding, the targeted attack can make the stable diffusion model generate an image that does not contain any dog-related content, although the prompt includes ``a dog".

However, the adversarial effect in\cite{zhuang2023pilot} is only limited to distorting the generation results. The work in\cite{zhang2024revealing} instead aims to mislead the generated images into a specific targeted category (e.g., cat) by appending certain words into the original text prompt. The added words do not directly relate to ``cat", but the final generated image will be a cat image. To do so, the adversarial prompt is learned via an objective that: (i) minimizes the classification loss of an image classifier, given that the input of the classifier is the image generated by the attacked DM, and its label is the target category (e.g., cat); (ii) bypasses a keyword detector that detects whether the adversarial prompt contains any word related to the target category; (iii) ensures that the similarity between the original prompt and the adversarial prompt is greater than a predefined threshold. This objective is optimized via gradient descent in the latent space, then the learned text embedding is converted back into the text space, resulting in the adversarial text prompt. Nevertheless, the adversarial prompt is unnatural in many cases, e.g., ``A combination lock secures the entrance to a secret chamber \textcolor{red}{fowl neuroscience}." The perturbation ``fowl neuroscience" can be detected easily by human inspection as its meaning does not relate to the original prompt. To make the adversarial prompt reasonable to human observers, the authors in\cite{liu2023discovering} first uses a LLM to generate $k$ candidate textual perturbations that can be appended to the text prompt without making the prompt semantically unnatural. Then, a loss function consisting of two different terms is optimized by gradient descent to choose the most suitable adversarial prompt from the $k$ candidates. The first term uses an image classifier as a robust discriminator to force the DM to generate mislabelled images. The second loss maximizes the similarity between the adversarial prompt and the nearest candidate prompt (in text embedding/latent space). As a result, an adversarial prompt ``A photo of a cat \textcolor{red}{tracking bears}" can make the DMs generate bear images instead of cat images, while the entire prompt is still semantically natural to human observers.

\begin{table*}[]
\centering
\caption{A comparison of learning objective between different adversarial attacks on DMs' input image.}
\label{table:adversarial-input-image}
\renewcommand{\arraystretch}{1.5} 
\begin{tabular}{|c|l|c|l|}
\hline
\textbf{Reference} & \multicolumn{1}{c|}{\textbf{Adversarial Learning Objective}} & \textbf{Space} & \multicolumn{1}{c|}{\textbf{Description}} \\ \hline
\multirow{2}{*}{PhotoGuard\cite{salman2023raising}} & $\min_\delta \| \mathcal{E}(c+\delta) - \mathcal{E}(c_{tgt})\|$ & Latent & Shift image representation towards a ``bad" target direction \\ \cline{2-4} 
 & $\min_\delta \| \mathcal{G}(c+\delta) - c_{tgt}\|$ & Image & Shift generated image towards a ``bad" target direction \\ \hline
Glaze\cite{shan2023glaze} & $\min_\delta \| \mathcal{E}(c+\delta) - \mathcal{E}(\mathcal{S}(c, T))\|$ & Latent & Shift image representation towards a ``wrong" image style \\ \hline
LDM-Robustness\cite{zhang2023robustness} & $\max_\delta \| \mathcal{E}(c+\delta) - \mathcal{E}(c)\|$ & Latent & Shift image representation away from itself \\ \hline
MFA\cite{yu2024step} & $\max_\delta \ex \| \mu(\epsilon_\theta(x_t,t,c+\delta) \|$ & Noise & Maximize the mean values of the predicted noise \\ \hline
\end{tabular}
\end{table*}

In the discussed above studies, the textual perturbation is often the entire words and it is appended to the text prompt or used to replace other words in the original prompt. There is another adversarial strategy investigated in\cite{kou2023character} that only replaces characters in some words of the text prompt instead of replacing the entire words. For example, given the prompt is ``A brown dog", changing from ``brown" to ``br0wn" may make the DMs generate a dog that is not brown. The replacement can be more stealthy by using characters with identical appearances to the ones being replaced. For instance, the Latin character o (U+006F) can be replaced by Cyrillic o (U+043E), and Greek o (U+03BF) with almost identical appearance but different encodings. Specifically, for each candidate replacement position, some visually similar characters are sampled as alternations for the original character at that position. These sampled characters are used to estimate the gradient of the optimal perturbation of images. Based on this gradient, the authors select a character that matches the best with the gradient direction. This is done via a learning objective that aims to maximize the adversarial effect while minimizing the different between the adversarial prompt and the original prompt to ensure stealthiness.

The main purpose of the presented above methods is to apply a small perturbation to the text prompt (e.g., appending or replacing several words/characters) to make the generated images misaligned with the provided text prompts although the two prompts are very similar to each other. On the other hand, the work\cite{liu2023riatig} takes an opposite approach: It tries to find an adversarial text prompt that is totally different from the original prompt, but leading to a similar generation result compared to the original prompt. This is formulated as an objective that maximizes the similarity between the generated benign image and the adversarial image, while the distance between the original prompt and the adversarial prompt (in latent/embedding space) is constrained to be less than a predefined threshold. This objective is achieved via a genetic-based optimization method that searches potential prompts with mutation. The process is similar to nature selection: Mutation is applied to generate different variants, increasing the diversity of population. The searching algorithm is repeated until finding out a text prompt satisfies the defined requirement. Especially, this attack is applicable for black-box settings, in which we do not have access to the DMs' parameters or any other information.

While distorting the generation results is a key application of prompt-based adversarial attacks, using this type of attack to bypass safety checking systems to generate sensitive NSFW contents is also an important research direction. Both studies \cite{qu2023unsafe} and \cite{chen2019detecting} investigated this topic and showed that NSFW safety checkers of public stable diffusion remain vulnerable to prompt-based adversarial attacks. However, the creation of adversarial prompts to bypass safety checkers is manual in these studies, resulting in a low bypass rate\cite{yang2024sneakyprompt}. Learning adversarial prompts is automatized in\cite{yang2023mma}, where the authors proposed a multi-modal adversarial attack to bypass safety checkers of public text-to-image DMs. In general, a safety checker often consists of two main components, a prompt-based safety filter and an image-based safety filter. The prompt filter verifies if the target prompt contains any sensitive, violent, offensive words like naked, nude, or zombie. To bypass the prompt filter, the authors in\cite{yang2023mma} combine gradient-driven optimization with greedy search to find out an adversarial prompt that has the most similar embedding values with the target offensive prompt, but do not contain any invalid words that are prohibited by the prompt filter. On the other hand, the image filter uses $M$ default NSFW embeddings, which convey sensitive/offensive content, to compare with the embedding/representation of the generated image. To bypass the image filter, the authors in\cite{yang2023mma} introduced a learning objective that minimize the similarity between the generated image's embedding and these $M$ NSFW embeddings. The final goal is to make the image's embedding different than all NSFW embeddings at a predefined threshold. As greedy search results in exhaustive resource consumption, the authors in\cite{yang2024sneakyprompt} deploys Reinforcement Learning (RL) to guide the token search process. This improves significantly the bypass rate compared to those applying baseline heuristic search like brute-force, greedy, and beam search.

Besides safety checkers, machine unlearning (MU)\cite{bourtoule2021machine} can also be used to protect DMs from sensitive content. MU is a process where a well-trained model is modified to forget specific data or information without having to be retrained from scratch\cite{cao2015towards, sekhari2021remember}. Thus, MU can be used to make DMs forget specific sensitive content, objects, styles, and cannot generate these types of content anymore. To attack such unlearned DMs, Zhang et al.\cite{zhang2023generate} proposed UnlearnDiffAtk, a prompt-based adversarial attack that can guide unlearned DMs towards generating sensitive contents. To do so, the authors prepare a target image that contains the desirable sensitive content or image style that has been unlearned by the DMs. Then, they use the noise-prediction loss of typical DMs to guide the gradient descent process of learning the textual perturbation. By doing so, UnlearnDiffAtk eliminates the need for an additional classifier or DM to guide the learning process.

In a real-world scenario, users may have several typos while writing prompts for conditional DMs. The authors in\cite{gao2023evaluating} consider these typos as potential adversarial perturbation, thus introducing three main types of real-world adversarial attacks based on greedy search: (i) Typo, which includes deleting, inserting, swapping, repeating, transforming case, adding space, and replacing characters (e.g., ``A \textcolor{red}{brownn} dog by the sea"); (ii) Glyph, which replaces certain characters with visually similar ones (e.g., ``A \textcolor{red}{br0wn} dog by the sea"); (iii) Phonetic, which replaces characters such that the perturbed words sound similar to the original words (e.g., ``A brown dog \textcolor{red}{buy} the \textcolor{red}{see}"). To exacerbate the adversarial effect, an objective function is proposed to maximize the distance between the image generated by the adversarial prompt and the image generated by the benign prompt, given that the distance between the two prompts is lower than a predefined threshold.

\subsubsection{Adversarial Attack on Fine-Tuning Images}
As personalization techniques like DreamBooth\cite{ruiz2023dreambooth} and Textual Inversion\cite{gal2022image} are used widely to fine-tune DMs, a special type of adversarial attack has emerged, focusing on perturbing such fine-tuning images. As described in Fig.~\ref{figure:adversarial}b, if adversarially perturbed images are used as fine-tuning images for personalization techniques, it will extract the personalized content to a feature map that is outside of the DMs' distribution. This makes the DMs generate low-quality or unrealistic images when the personalized content is referred to in the text prompt. Thanks to this effect, this type of adversarial attack is often used with positive purposes like copyright projection and watermark embedding. For instance, artists can add adversarial perturbation into their arts. Then, if the protected arts are used to fine-tune DMs, the DMs will not be able to generate high-quality images that mimic the artists' style/content.

AdvDM\cite{liang2023adversarial} is among the first studies investigated this type of adversarial attack. The mentioned perturbation $\delta$ is learnt by minimizing $p_\theta(x+\delta)$, with $\delta$ is constrained to be less than a small constant to retain the quality of the perturbed images. This objective is opposite to the training objective of benign DMs, where $p_\theta(x)$ is maximized. Intuitively, while the objective of DMs is to maximize the probability of generating images from the training data's distribution, AdvDM aims to reduce this probability as much as possible to degrade the generation quality. In practice, $p_\theta(x+\delta)$ is estimated by Monte Carlo, while Textual Inversion is used as the main personalization technique for evaluation. Liang et al.\cite{liang2023mist} developed AdvDM further by introducing Mist, a combination of AdvDM\cite{liang2023adversarial} and PhotoGuard\cite{salman2023raising}. Specifically, the perturbation learning objective of Mist is a fuse of the two learning objectives in AdvDM (i.e., semantic loss) and PhotoGuard (textual loss), making the adversarial more robust and efficient. Furthermore, Mist is applicable for various DM-based method such as Textual Inversion, DreamBooth, and image-to-image generation.

Anti-DreamBooth is introduced by Van et al. in another work\cite{van2023anti}. In this paper, the authors aim to disturb DreamBooth by a tiny perturbation added into the fine-tuning images. This perturbation is learned via a bi-level optimization problem that maximizes the conditional loss of the targeted DM with regard to the perturbation $\delta$, given that the DM is concurrently fine-tuned by DreamBooth on the perturbed images.

However, the main goal of the presented methods is only limited to degrading the quality of generated images under personalization techniques. The authors in\cite{zhu2024watermark} take a step further by embedding watermarks into the generated images. The aim is to cause the DMs fine-tuned by perturbed images to generate low-quality images with a visible predefined watermark. This is realized via a combination of three different loss functions: (i) an adversarial loss that minimizes the distance between the generated adversarial images and the original image (in latent space); (ii) a GAN loss that minimizes the difference between the perturbed image and the original image to ensuring stealthiness; and (iii) a perturbation loss that limits the scale of the perturbation, focusing on only image region that contains the watermark. This seems very similar to a backdoor attack if we consider the adversarial perturbation as backdoor trigger, and the watermark as backdoor target. However, the fundamental difference is that backdoor attacks manipulate deeply the DMs' parameters, architecture, and forward/reverse processes, while adversarial attacks only focus on adding a small perturbation into the images or text prompt without having access to the models.

\subsection{Membership Inference Attacks}
In this study, based on the capabilities of attackers, MIAs are divided into three categories: i) black-box attack, ii) gray-box attack, and iii) white-box attack, as described in Table (\ref{table:summary_of_MIAattacks})

\begin{table*}[]
\centering
\caption{Summary of membership inference attacks on Diffusion Models. Internal values are model's internal information such as loss and gradient. B-box, G-box, and W-box indicate black-box, gray-box, and white-box, respectively. Cond. and Uncond. denote conditional and unconditional DMs, and *  indicates proposed datasets.}
\label{table:summary_of_MIAattacks}
\resizebox{\textwidth}{!}{%
\renewcommand{\arraystretch}{1.2}
\begin{tabular}{|c|cc|ccc|c|c|c|c|}
\hline
\rowcolor[HTML]{EFEFEF} 
\cellcolor[HTML]{EFEFEF} &
  \multicolumn{2}{c|}{\cellcolor[HTML]{EFEFEF}\textbf{Target Model}} &
  \multicolumn{3}{c|}{\cellcolor[HTML]{EFEFEF}\textbf{Attack Setting}} &
  \cellcolor[HTML]{EFEFEF} &
  \cellcolor[HTML]{EFEFEF} &
  \cellcolor[HTML]{EFEFEF} &
  \cellcolor[HTML]{EFEFEF} \\ \cline{2-6}
\rowcolor[HTML]{EFEFEF} 
\multirow{-2}{*}{\cellcolor[HTML]{EFEFEF}\textbf{Ref.}} &
  \multicolumn{1}{c|}{\cellcolor[HTML]{EFEFEF}\textbf{Uncond.}} &
  \textbf{Cond.} &
  \multicolumn{1}{c|}{\cellcolor[HTML]{EFEFEF}\textbf{B-box}} &
  \multicolumn{1}{c|}{\cellcolor[HTML]{EFEFEF}\textbf{G-box}} &
  \textbf{W-box} &
  \multirow{-2}{*}{\cellcolor[HTML]{EFEFEF}\textbf{Attack Access}} &
  \multirow{-2}{*}{\cellcolor[HTML]{EFEFEF}\textbf{Attack Feature}} &
  \multirow{-2}{*}{\cellcolor[HTML]{EFEFEF}\textbf{Attack Method}} &
  \multirow{-2}{*}{\cellcolor[HTML]{EFEFEF}\textbf{Datasets}} \\ \hline
\cite{wu2022membership} &
  \multicolumn{1}{c|}{} &
  \checkmark &
  \multicolumn{1}{c|}{\checkmark} &
  \multicolumn{1}{c|}{} &
   &
  Final output image &
  Generated image &
  Classifier &
  \begin{tabular}[c]{@{}c@{}}MSCOCO, VG\\ Laion-400M, CC3M\end{tabular} \\ \hline
\cite{zhang2024generated} &
  \multicolumn{1}{c|}{\checkmark} &
  \checkmark &
  \multicolumn{1}{c|}{\checkmark} &
  \multicolumn{1}{c|}{} &
   &
  Final output image &
  Generated image &
  Classifier &
  \begin{tabular}[c]{@{}c@{}}CIFAR-10, CelebA, Laion\\ CelebA-HQ, ImageNet\end{tabular} \\ \hline
\cite{pang2023black} &
  \multicolumn{1}{c|}{} &
  \checkmark &
  \multicolumn{1}{c|}{\checkmark} &
  \multicolumn{1}{c|}{} &
   &
  Final output image &
  Generated image &
  \begin{tabular}[c]{@{}c@{}}Threshold, Classifier\\ Likelihood ratio\end{tabular} &
  \begin{tabular}[c]{@{}c@{}}CelebA-Dialog\\ WIT, MSCOCO\end{tabular} \\ \hline
\cite{li2024towards} &
  \multicolumn{1}{c|}{\checkmark} &
  \checkmark &
  \multicolumn{1}{c|}{\checkmark} &
  \multicolumn{1}{c|}{} &
   &
  Final output image &
  Generated image &
  Classifier &
  \begin{tabular}[c]{@{}c@{}}CIFAR-10. STL10-U\\ Laion-5B\\ Laion-by-DALL-E*\end{tabular} \\ \hline
\cite{duan2023diffusion} &
  \multicolumn{1}{c|}{\checkmark} &
  \checkmark &
  \multicolumn{1}{c|}{} &
  \multicolumn{1}{c|}{\checkmark} &
   &
  \begin{tabular}[c]{@{}c@{}}Noise schedule\\ Intermediate outputs\end{tabular} &
  \begin{tabular}[c]{@{}c@{}}Posterior estimation\\ error\end{tabular} &
  \begin{tabular}[c]{@{}c@{}}Classifier\\ Threshold\end{tabular} &
  \begin{tabular}[c]{@{}c@{}}CIFAR-10/100, STL10-U\\ Tiny-ImageNet, Pokemon\\ COCO2017-val, Laion-5B\end{tabular} \\ \hline
\cite{tang2023membership} &
  \multicolumn{1}{c|}{\checkmark} &
   &
  \multicolumn{1}{c|}{} &
  \multicolumn{1}{c|}{\checkmark} &
   &
  \begin{tabular}[c]{@{}c@{}}Noise schedule\\ Intermediate outputs\end{tabular} &
  \begin{tabular}[c]{@{}c@{}}Posterior estimation\\ error\end{tabular} &
  Quantile regression &
  \begin{tabular}[c]{@{}c@{}}CIFAR-10/100, STL100\\ Tiny-ImageNet\end{tabular} \\ \hline
\cite{kong2023efficient} &
  \multicolumn{1}{c|}{\checkmark} &
  \checkmark &
  \multicolumn{1}{c|}{} &
  \multicolumn{1}{c|}{\checkmark} &
   &
  \begin{tabular}[c]{@{}c@{}}Noise schedule\\ Intermediate outputs\end{tabular} &
  \begin{tabular}[c]{@{}c@{}}Posterior estimation\\ error\end{tabular} &
  Threshold &
  \begin{tabular}[c]{@{}c@{}}CIFAR-10/100, Tiny-ImageNet\\ COCO2017, Laion-5B\end{tabular} \\ \hline
\cite{zhai2024membership} &
  \multicolumn{1}{c|}{} &
  \checkmark &
  \multicolumn{1}{c|}{} &
  \multicolumn{1}{c|}{\checkmark} &
   &
  \begin{tabular}[c]{@{}c@{}}Noise schedule\\ Intermediate outputs\end{tabular} &
  Likelihood &
  \begin{tabular}[c]{@{}c@{}}Classifier\\ Threshold\end{tabular} &
  \begin{tabular}[c]{@{}c@{}}Pokemonn Flickr\\ MSCOCO, Laion\end{tabular} \\ \hline
\cite{hu2023membership} &
  \multicolumn{1}{c|}{\checkmark} &
   &
  \multicolumn{1}{c|}{} &
  \multicolumn{1}{c|}{} &
  \checkmark &
  \begin{tabular}[c]{@{}c@{}}Noise schedule\\ Internal values\end{tabular} &
  \begin{tabular}[c]{@{}c@{}}Loss\\ Likelihood\end{tabular} &
  Threshold &
  FFHQ, DRD \\ \hline
\cite{carlini2023extracting} &
  \multicolumn{1}{c|}{\checkmark} &
   &
  \multicolumn{1}{c|}{} &
  \multicolumn{1}{c|}{} &
  \checkmark &
  \begin{tabular}[c]{@{}c@{}}Noise schedule\\ Internal values\end{tabular} &
  Loss &
  Likelihood ratio &
  CIFAR-10 \\ \hline
\cite{matsumoto2023membership} &
  \multicolumn{1}{c|}{\checkmark} &
   &
  \multicolumn{1}{c|}{\checkmark} &
  \multicolumn{1}{c|}{} &
  \checkmark &
  \begin{tabular}[c]{@{}c@{}}Final output image\\ Noise schedule\\ Internal values\end{tabular} &
  Loss &
  Threshold &
  \begin{tabular}[c]{@{}c@{}}CIFAR-10\\ CelebA\end{tabular} \\ \hline
\cite{dubinski2024towards} &
  \multicolumn{1}{c|}{} &
  \checkmark &
  \multicolumn{1}{c|}{} &
  \multicolumn{1}{c|}{} &
  \checkmark &
  \begin{tabular}[c]{@{}c@{}}Noise schedule\\ Internal values\end{tabular} &
  Loss &
  Threshold &
  \begin{tabular}[c]{@{}c@{}}Pokemon\\ Laion-mi*\end{tabular} \\ \hline
\cite{pang2023white} &
  \multicolumn{1}{c|}{\checkmark} &
  \checkmark &
  \multicolumn{1}{c|}{} &
  \multicolumn{1}{c|}{} &
  \checkmark &
  \begin{tabular}[c]{@{}c@{}}Noise schedule\\ Internal values\end{tabular} &
  Gradient &
  Classifier &
  \begin{tabular}[c]{@{}c@{}}CIFAR-10\\ MSCOCO, ImageNe\end{tabular} \\ \hline
\end{tabular}%
}
\end{table*}

\subsubsection{Black-box MIA}
In this scenario, Matsumoto et al.\cite{matsumoto2023membership} directly adopt the full black-box attack of GAN-Leaks\cite{chen2020gan} for DDIM\cite{song2020denoising} and evaluate how vulnerable DMs are compared to deep convolutional GAN (DCGAN) \cite{radford2015unsupervised}. Specifically, given a candidate image, GAN-Leaks\cite{chen2020gan} generates 12,000 images to form a generator's output space and then infers the membership based on the smallest reconstruction error. However, since DMs are more complex, this method causes a huge bottleneck in terms of computational resources and time when sampling such a large number of samples. Additionally, due to the difference in training and inference procedures, the results demonstrate that existing approaches designed for GAN are significantly less effective when applied to DMs.

Concurrently, Wu et al.\cite{wu2022membership} analyze privacy risks of state-of-the-art text conditional generation models when an arbitrary adversary attempts to determine whether a candidate image belongs to the target model's training dataset without having its ground-truth caption. In this work, the attackers are assumed to have a small subset of the target model's training data, as well as a small set of local non-member data to form an auxiliary dataset to train attack models.
When the adversaries try to attack a target model, they first leverage a third-party image captioning tool (e.g., BLIP\cite{li2022blip}) to generate a caption for the candidate image, which is then fed into the target model to obtain the generated image. Subsequently, different attack methods, including pixel-level and semantic-level attacks, are designed based on various perspectives, such as the quality of the generated image, the reconstruction error between the candidate image and the generated image, and the faithful semantic reflection of the textual prompt in the generated image. As a result, when evaluating the proposed methods on an LDM\cite{rombach2022high} pre-trained with the LAION-400M\cite{schuhmann2021laion} dataset, the auxiliary dataset is constructed by selecting human face subsets from the LAION-Face \cite{zheng2022general} dataset as members and from the MS COCO Face\cite{chen2015microsoft}  and Visual Genome Face \cite{krishna2017visual} datasets as non-members. Notably, the best approach achieves a remarkable performance of over 97\% accuracy on the non-member dataset. 
However, these proposed approaches\cite{wu2022membership} require access to a small subset of the target model's training data, which is not always feasible in most realistic scenarios. 

In another work, Zhang et al.\cite{zhang2024generated} argue that the distribution of images generated from the target model will exhibit, to some extent, an approximation of the training data distribution. Thus, under the black-box setting, attackers can simply query the target generative model via an API to produce images and use them as positive samples, while drawing negative samples from an auxiliary dataset that has a similar domain as the target model's training data. These negative and positive samples are then used to train a ResNet18-based\cite{he2016deep} classifier for membership inference, which is then evaluated on the target model's training data and the negative samples. Since this MIA method is both architecture-agnostic and task-agnostic, it can be applied to various generative models, such as GANs, DMs, VAEs, and IFs. As a result, the proposed method achieves remarkable performance with AUCs over 0.99 on DDPM\cite{ho2020denoising}, DDIM\cite{song2020denoising}, and Fast-DPM\cite{kong2021fast} trained on CIFAR-10\cite{krizhevsky2009learning} or CelebA\cite{liu2015deep}; AUCs over 0.9 on VQGAN\cite{esser2021taming}, text-conditional LDM\cite{rombach2022high}, and LIIF\cite{chen2021learning} trained on ImageNet\cite{deng2009imagenet}, Laion-400M\cite{schuhmann2021laion}, and CelebA-HQ\cite{karras2017progressive}, respectively.
Nonetheless, obtaining positive samples requires numerous queries to the target model, which is often restricted by the API providers. In addition, since the membership classifier is trained by choosing an auxiliary dataset on behalf of all non-member samples and using the model's generated image distribution to approximate the actual member distribution, the attack model can be overfitted, and the generalization of the membership boundary is questionable.

Instead of directly developing attack methods on the target model, Pang et al.\cite{pang2023black} propose utilizing an auxiliary dataset to train a shadow model that mimics the behavior of the target model. Subsequently, this shadow model is employed to generate data for training the attack models. 
% Additionally, the authors direct their consideration to the data privacy issues that arise during the fine-tuning of a pre-trained large-scale model for downstream tasks. Specifically, given the full open-source and intensive availability of models pre-trained on large-scale datasets, the pre-training fine-tuning paradigm has become increasingly popular and raises concern about the privacy of fine-tuning datasets.
Derived directly from the objective function of DMs, the authors demonstrate that the similarities between the generated image and the query image of the member samples are significantly higher compared to the non-member samples. Therefore, given a query image, the attacker first utilizes its caption, either the actual caption or one generated by a third-party captioning tool (e.g., BLIP2\cite{li2023blip}), to query the target model and obtain the generated image. Following this, the attacker employs an image encoder (e.g., DETR\cite{carion2020end}, BEiT\cite{bao2021beit}, DeiT\cite{touvron2021training}, etc.) to extract the features of both the query and generated images, then compute their semantic similarity score based on these embedding vectors. This similarity is then used to determine the membership by applying various methods, including threshold comparison, distribution-based analysis, and MLP classifiers.

Concurrently, Li et al.\cite{li2024towards} analyze the privacy of DDIM and stable diffusion \cite{rombach2022high} when attackers utilize the variation API, i.e., the image-to-image procedure, to alter the candidate image and compare it with the original, subsequently inferring the membership. Specifically, when querying a DDIM with a candidate image $x$, the attacker first diffuses it into a noisy image by randomly incorporating $t$-step Gaussian noise $\epsilon_0 \sim \mathcal{N}(0,\mathbf{I})$ following (\ref{add_noise_ddpm}). Then, the target model's sampling process is employed to denoise the noisy image to obtain the reconstructed image $\hat{x}_i$. By performing the variation API $n$ independent times to obtain $n$ reconstructed images $\{\hat{x}_{1}, \hat{x}_{2}, ..., \hat{x}_{n}\}$, subsequently the final reconstructed image is computed by taking average: $\hat{x} = \frac{1}{n}(\hat{x}_{1} + \hat{x}_{2} + ... + \hat{x}_{n})$.
Intuitively, averaging over multiple independent reconstructed images $\hat{x}_i$ would greatly reduce the reconstruction error, make it distinguishable from the non-member samples. Exploiting this, the authors first train a ResNet18-based\cite{he2016deep} binary classifier $f_{R}$ as a proxy task based on the reconstruction error: $v = ||\hat{x} - x||$. Then, the distance between the query image $x$ and the variational image $\hat{x}$ is computed by taking the negative value of probability that sample is predicted as a member: $D(x, \hat{x}) = -f_R(v)$. Finally, the membership is determined by comparing the distance with a certain threshold.
% When applying to stable diffusion, a text-conditional diffusion model, the caption used for the sampling process could be either the ground truth text or the one generated through BLIP2\cite{li2023blip}. 
% In addition, since stable diffusion models produce larger images, directly comparing the variational image with the original leads to less effective performance. Therefore, the authors propose querying the variation API twice and measuring the distance between two variational images using the SSIM metric \cite{wang2004image} instead.
Although this method can work in a black-box setting, similar to \cite{wu2022membership}, it still requires access to a subset of the target model's actual training data, which is not realistic in most scenarios.

\subsubsection{Gray-box MIA}
Under the gray-box setting, Duan et al. \cite{duan2023diffusion} introduce SecMI, which aims to infer membership of both DDPM and stable diffusion models.
Assuming that the member and non-member sets share the same distribution, SecMI assesses the posterior estimation with deterministic sampling and reversing processes to infer the membership. Specifically, at timestep $t$, DMs try to approximately estimate the reverse process with the forward process by minimizing the following loss function: 
\begin{equation}
\label{MIAequation48}
    \ell_t = E_q \Biggl[ \frac{1}{2\sigma^2_t}|| \widetilde{\mu_t}(x_t, x_0) - \mu_\theta(x_t, t) || \Bigg]\ ,
\end{equation}
where $\widetilde{\mu_t}(x_t, x_0)$ is the mean of posterior distribution $q(x_{t-1}|x_t, x_0)$ (i.e., diffusion process), $\mu_\theta(x_t, t)$ refers to the estimated mean for $p_\theta(\hat{x}_{t-1}|x_t)$ (i.e., denoising process), and $\theta$ denotes the weights of the DM. Then, the local estimation error of a single data point $x_0$ at timestep $t$ can be approximated as:
\begin{equation}
\label{MIAequation48_}
    \ell_{t, x_0} = ||\hat{x}_{t-1} - x_{t-1}||^2,
\end{equation}
where $\hat{x}_{t-1} \sim p_\theta(\hat{x}_{t-1}|x_t)$ and $x_{t-1} \sim q(x_{t-1}|x_t, x_0)$.\\ 
In addition, by leveraging deterministic properties of DDIM\cite{song2020denoising},  the authors define two deterministic processes:
\begin{equation}
    \label{MIAequation49}
    x_{t+1} = \phi_\theta(x_t, t) = \sqrt{\Bar{\alpha}_{t+1}}f_\theta(x_t, t) + 
    \sqrt{1 - \Bar{\alpha}_{t+1}}\epsilon_\theta(x_t, t)
\end{equation}
\begin{equation}
    \label{MIAequation50}
    x_{t-1} = \psi_\theta(x_t, t) \\
    = \sqrt{\Bar{\alpha}_{t-1}}f_\theta(x_t, t) + 
    \sqrt{1 - \Bar{\alpha}_{t-1}}\epsilon_\theta(x_t, t),
\end{equation}
where $f_\theta(x_t, t) = \frac{x_t - \sqrt{1 - \Bar{\alpha}_t}\epsilon_\theta(x_t, t)}{\sqrt{\Bar{\alpha}_t}}.$
Subsequently, for a given sample $x_0$, at timestep $t$, the approximated posterior estimation error is computed following (\ref{MIAequation48_}), (\ref{MIAequation49}), (\ref{MIAequation50}) and defined as \textit{t-error}:
\begin{equation}
    \label{MIAequation52}
    \widetilde{\ell}_{t, x_0} = ||\psi_\theta(
    \phi_\theta(
    \widetilde{x}_t, t), t) - \widetilde{x}_{t}|| ^2,
\end{equation}
where $\widetilde{x}_{t} = \phi_\theta(\cdots\phi_\theta(\phi_\theta(x_0, 0), 1), t-1)$ is the deterministic reverse process.
After computing estimation error $\widetilde{\ell}_{t_{SEC}, x_0}$ at a empirically selected timestep $t_{SEC}$ given a sample $x_0$, membership can be inferred by comparing this error with a threshold or employing a binary classifier.

In another work, Tang et al.\cite{tang2023membership} directly adopt \textit{t-error} from \cite{duan2023diffusion} following (\ref{MIAequation52}) and learn a quantile regression model $q_\alpha(x)$ that predicts the $\alpha$-quantile of the \textit{t-error} $\widetilde{\ell}_{t, x}$ for each sample $x$ in the auxiliary dataset \textit{D}, where $\alpha$ is a parameter controlling the false positive rate. As a result, instead of applying a uniform threshold across all samples, \cite{tang2023membership} employs $q_\alpha$ to predict the sample-conditioned $\alpha$-quantile of the \textit{t-error} for each sample and uses it as the fine-grained per-sample threshold to classify the membership status.

Having a similar idea of leveraging the deterministic properties of DDIM\cite{song2020denoising}, Kong et al.\cite{kong2023efficient} propose PIA, which can significantly reduce the number of required queries compared to SecMI\cite{duan2023diffusion}. Specifically, in the DDIM framework, if standard deviation $\sigma_t$ = 0, the process of adding noise becomes deterministic. Thus, given two points $x_0$ and $x_k$, we can determine any other points $x_t$, which is called \textit{ground-truth trajectory} and obtained from $x_0$ and $x_k$:
\begin{equation}
\label{MIAequation54}
    x_t = \sqrt{\Bar{\alpha}_{t}}x_0 + \sqrt{1-\Bar{\alpha}_{t}}\cdot{\Bar{\epsilon}_k},
\end{equation}
where ${\Bar{\epsilon}_k} = \frac{x_k-\sqrt{{\Bar\alpha}_k}x_0}{\sqrt{1-\Bar{\alpha}_k}}$
is the noise added to $x_0$ to obtain $x_k$.
The authors choose $k=0$ and approximately compute the noise $\Bar{\epsilon}_0$ initialized at $t=0$ following:
\begin{equation}
\label{MIAequation56}
    \Bar{\epsilon}_0 \approx \epsilon_\theta(\sqrt{\Bar{\alpha}_{0}}x_0 + \sqrt{1-\Bar{\alpha}_{0}}{\Bar{\epsilon}_0}, 0) \approx \epsilon_\theta(x_0, 0).
\end{equation}
Subsequently, following Eq.(\ref{MIAequation54}) and Eq.(\ref{MIAequation56}), we can obtain the ground-truth points $x_t$ and $x_{t-t'}$, and sample the predicted point $x'_{t-t'}$  from $x_t$ following the deterministic sampling process of DDIM\cite{song2020denoising} (\ref{equation26}) when $\sigma_t$ = 0. Following this, the distance between $x_{t-t'}$ and $x'_{t-t'}$ is computed using $l_p-\text{norm}$ and ends up with the simple form:
\begin{align}
\label{MIAequation57}
    R_{t,p} &= || \Bar{\epsilon}_0 - \epsilon_{\theta}(\sqrt{\Bar{\alpha}_t}x_0 + \sqrt{1 - \Bar{\alpha}_t}\Bar{\epsilon}_0,t)||_p \\
    &\approx || \epsilon_\theta(x_0, 0) - \epsilon_{\theta}(\sqrt{\Bar{\alpha}_t}x_0 + \sqrt{1 - \Bar{\alpha}_t}\epsilon_{\theta}(x_0, 0),t)||_p.
\end{align}
Finally, the membership is inferred by comparing $R_{t,p}$ with a threshold $\tau$. 
% Additionally, Kong et al.\cite{kong2023efficient} also propose PIAN (PIA Normalized) which normalizes $\Bar{\epsilon}_0$ to improve the performance.

% Furthermore, this method can be applied to continuous-time diffusion models when $\sigma_t$ = 0 and the reverse process (\ref{equation32}) becomes an ordinary differential equation (ODE). Specifically, approximating $||x_{t-t'} - x'_{t-t'}|| \approx dx_t = (f_t(x_t) - \frac{1}{2}g^2_t s_\theta(x_t ,t))dt$ and ignoring $dt$, the attack metric becomes:
% \begin{equation}
% \label{MIAequation59}
%     R_{t,p} = || f_t(x_t) - \frac{1}{2}g^2_t s_\theta(x_t ,t)||_p
% \end{equation}

Concurrently, Zhai et al.\cite{zhai2024membership} propose a gray-box method targeting text-conditional DMs, utilizing conditional likelihood discrepancy (CLiD). Specifically, recall that the training objective of DMs is to estimate the reverse process matching with the forward process, i.e., minimize $\mathcal{L}(x_0) = \ex_q \left[ D_\text{KL}(q(x_{t-1}|x_t,x_0)||p_\theta(x_{t-1}|x_t)) \right]$, the overfitting phenomenon of DMs tends to result in lower estimation errors for member samples $x_m$ compared to non-member samples $x_n$: $\mathcal{L}(x_{m}) \leq \mathcal{L}(x_{n})$.
This difference becomes more salient when it comes to text-conditional DMs, where the denoising process is conditioned by the caption $c$:
\begin{equation}
    \label{MIAequation61}
    \mathcal{L}(x_{n}, c_{null}) - \mathcal{L}(x_{m}, c_{null})
    \leq \mathcal{L}(x_{n}, c) - \mathcal{L}(x_{m}, c),
\end{equation}
where $c_{null}$ denotes an empty text input to the text-conditional DM.
Given $I(x, c)$ is the conditional likelihood discrepancy of image-text data point $(x, c)$: $I(x, c) = \mathcal{L}(x, c_{null}) - \mathcal{L}(x, c)$,
following (\ref{MIAequation61}), we have:
\begin{equation}
    \label{MIAequation63}
    I(x_m, c) \geq I(x_n, c).
\end{equation}
And $I(x, c)$ can be expressed as:
\begin{align}
    I(x, c) &= \ex_{t, \epsilon} \left[||\epsilon_\theta(x_t, t, c_{null}) - \epsilon||^2\right]\\
      &   - \ex_{t, \epsilon} \left[||\epsilon_\theta(x_t, t, c) - \epsilon||^2\right]
\end{align}
Consequently, Zhai et al.\cite{zhai2024membership} estimate $I(x, c)$ and $\mathcal{L}(x, c)$ by performing Monte Carlo estimation, then combining them to obtain the attack feature and utilizing threshold-based attack and XGBoost-based\cite{chen2016xgboost} classifier to infer the membership.

\subsubsection{White-box MIA}
Leveraging the overfitting tendency in deep neural networks, most MIAs follow the common assumption that samples from the training dataset exhibit the lower loss compared to non-member samples.\\ Following this, while Hu et al.\cite{hu2023membership} and Matsumoto et al.\cite{matsumoto2023membership} directly utilize DM's loss at each timestep for threshold-based membership inference attacks, Dubiński et al.\cite{dubinski2024towards} focus on modifying the diffusion process to extract the model's loss information at various perspectives to enhance their threshold-based and classifier-based attacks' performances. In addition, \cite{hu2023membership} also proposes computing log-likelihood of the candidate image following \cite{song2020score} as a different attack feature for the threshold-based approach.

On the other hand, Carlini et al. \cite{carlini2022membership} argue that while threshold-based attacks are effective for non-membership inference, they lack precision for classifying member samples, which is considerably more important. Therefore, the authors propose adopting the Likelihood Ratio Attack (LiRA) from \cite{carlini2022membership}, which analyses the distribution of the losses to classify the membership, assuming that the members' losses and non-members' losses belong to two different distributions. 
To this end, LiRA first trains multiple shadow models $\{f_1, f_2, ..., f_{i}, ...\}$, each on a random subset of the training dataset, to imitate the behavior of the target model. Then, for a candidate sample $x$, the loss $\mathcal{L}(x; f_i)$ obtained from the shadow model $f_i$ is assigned to one of two sets: the set of member losses if $f_i$ did see $x$ during training, and the set of non-member losses if $x$ was not in $f_i$'s training data. Following this, the member loss distribution $D_{m}$ and non-member loss distribution $D_{n}$ are established by fitting Gaussians to the member losses and non-member losses sets. Finally, to predict the membership of the sample $x$ with the target model $f^*$, it computes the loss $l = \mathcal{L}(x; f^*)$ and then measures whether $Pr(l|D_{m}) > Pr(l|D_{n})$, i.e. which distribution the loss of target sample is more likely to belong to. As a result, by training 16 class-conditional DMs on CIFAR-10 as the shadow models and computing the losses at timestep $t=100$, Carlini et al.\cite{carlini2023extracting} achieve a TPR of over 70\% at a FPR of just 1\%. In contrast, when applied to state-of-the-art classifiers, LiRA achieves less than 20\% TPR at the same FPR. This demonstrates that DMs are significantly less private than classification models trained on the same data.
Although Carlini et al.\cite{carlini2023extracting} employ a more insightful method to exploit the loss feature, this approach necessitates training multiple shadow models and retraining when obtaining new target data points. This process is highly computationally intensive, particularly for large-scale DMs.

While relying on the loss information is an intuitive attacking approach, Pang et al.\cite{pang2023white} propose a new framework called GSA that can be applied to both unconditional and text-conditional DMs. This approach brings up a novel perspective by leveraging the gradient information as the attack feature, which might reflect more insightfully how the target model responds to member and non-member samples.
However, large-scale models with a huge number of parameters, such as DMs, often yield extremely high-dimensional gradient information. Extracting this feature effectively and efficiently is still a challenging task. To this end, the authors propose two algorithms, $GSA_1$ and $GSA_2$, utilizing sub-sampling and aggregation techniques to reduce dimensionality and retain important information. Specifically, for $K$ steps uniformly sub-sampled from the total diffusion steps $T$, $GSA_1$ computes loss value $\mathcal{L}_t$ for each timestep $t$, then takes the mean value of these losses:
\begin{equation}
    \Bar{\mathcal{L}} = \frac{1}{|K|}\sum_{t \in K}{\mathcal{L}_t}
\end{equation}
Following this, $\Bar{\mathcal{L}}$ is utilized to perform back-propagation on the target model with $N$ layers of parameters, and produce gradients for each layer. Then, $GSA_1$ performs $l_2-$norm on the gradient information of each layer to obtain their representative values, ultimately yielding the gradient vector $G$ to train a machine learning model for membership inference: 
\begin{equation}
    G \leftarrow \Biggl[ \Bigl|\Bigl|\frac{\partial\Bar{\mathcal{L}}}{\partial{W_1}}\Bigl|\Bigl|^2_2, \Bigl|\Bigl|\frac{\partial\Bar{\mathcal{L}}}{\partial{W_2}}\Bigl|\Bigl|^2_2, ...,   \Bigl|\Bigl|\frac{\partial\Bar{\mathcal{L}}}{\partial{W_N}}\Bigl|\Bigl|^2_2 \Biggl].
\end{equation}
While $GSA_1$ chooses to take the average of the losses before performing the back-propagation to reduce the computational overhead and accept the loss of information, $GSA_2$ opts to perform the back-propagation for the loss $\mathcal{L}_t$ at each timestep $t$ to obtain the gradient vector $G_t$, then compute the average of them to output the final attack vector: 
\begin{equation}
    G = \frac{1}{|K|}\sum_{t \in K}{G_t}.
\end{equation}
As a result, $GSA_2$ retains more useful information and achieves better performance, yet requires extremely high computational costs.

\section{Defenses for Diffusion Models}\label{section:defense-DM}
It has been shown in section~\ref{section:attack-DM} that there are a variety of attacks threatening DMs. Thus, investigating possible countermeasures for these attacks is obviously an important topic. This section provides a comprehensive survey of state-of-the-art defense methods for DMs against the presented attacks.

\subsection{Countermeasures of Backdoor Attacks}\label{subsection:countermeasure-backdoor}
In practice, it is especially challenging to detect backdoor attacks targeting on DMs due to its stealthiness: (i) Backdoored DMs still perform normally on benign inputs, and (ii) the input data space (e.g., image, text, audio) is usually too large for searching-based methods that try to identify the trigger. Furthermore, the processes of DMs are totally different from standard ML models like regression or classification models, making existing backdoor detection methods relying on labeled data\cite{li2021neural, liu2018fine, chen2019detecting} inapplicable for DMs.

\begin{figure}[h!]
	\centering
	\includegraphics[scale=0.16]{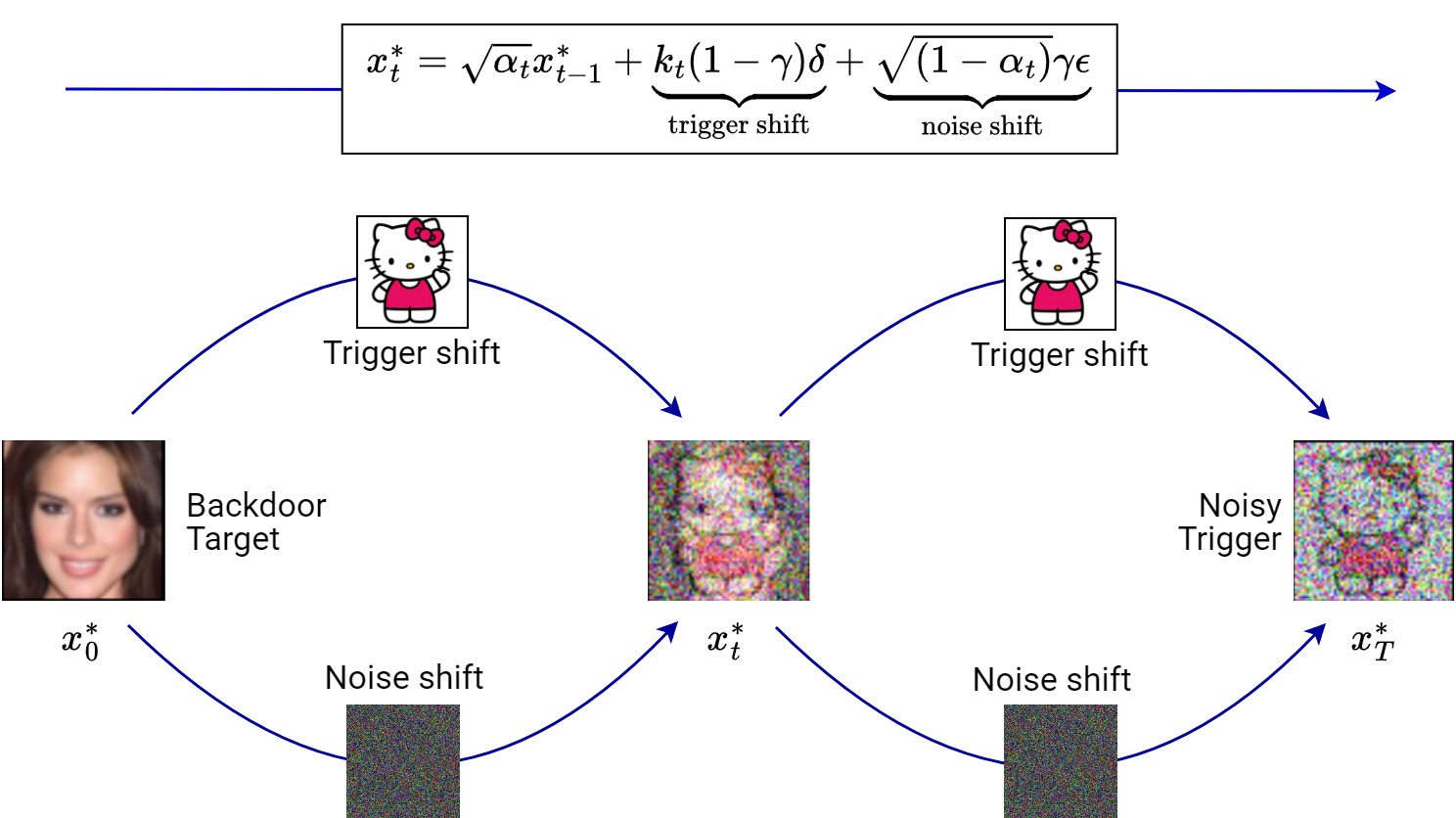}
	\caption{A visualization of the backdoored diffusion process in TrojDiff\cite{chen2023trojdiff}, from the perspective of distribution shift.}%\cite{BOOK:1}}
	\label{figure:trojdiff}
\end{figure}

Fortunately, the diffusion and reverse processes of DMs do have certain special properties to be exploited for detecting the embedded backdoor patterns\cite{an2024elijah, an2023remove}. The distribution shift caused by the backdoor trigger can be observed clearly from the presented TrojDiff and BadDiffusion methods. In the diffusion process of each work, the data distribution is shifted gradually from the target image $x_0^*$ towards the noisy trigger $x_T^*$. For instance, the forward transition of TrojDiff presented in (\ref{equation:trojdiff2}) can be reparameterized into the following form:
\begin{equation}
    x^*_t = \sqrt{\alpha_t}x^*_{t-1} + \underbrace{k_t(1-\gamma)\delta}_\textrm{trigger shift} + \underbrace{\sqrt{(1-\alpha_t)}\gamma\epsilon}_\textrm{noise shift},
\end{equation}
where $\epsilon \in \mathcal{N}(0,\mathbf{I})$, and $\delta$ is the backdoor trigger (i.e., the trigger distribution). As shown in Fig.~\ref{figure:trojdiff}, the ``trigger shift" term represents a small amount of distribution shift that guides the data distribution towards the trigger distribution $\delta$ (i.e., the hello-kitty image in Fig.~\ref{figure:trojdiff}) in each diffusion step. The scale/amount of this trigger shift is $k_t(1-\gamma)$. Concurrently, the ``noise shift" makes the data gradually resembles a Gaussian noise. The combination of both these shift terms results in the noisy trigger $x_T^*$ at the final diffusion step (i.e., the noisy hello-kitty image in Fig.~\ref{figure:trojdiff}). This observation is also true for BadDiffusion when we apply reparameterization on BadDiffusion's forward transition (\ref{equation:baddifussion3}):
\begin{equation}
    x^*_t = \sqrt{\alpha_t}x^*_{t-1} + \underbrace{(1-\sqrt{\alpha_t})x_s^\delta}_\textrm{trigger shift} + \underbrace{\sqrt{(1-\alpha_t)}\epsilon}_\textrm{noise shift},
\end{equation}
where $x_s^\delta$ is the noisy trigger image. In this case, the scale of the trigger shift is $(1-\sqrt{\alpha_t})$.

Intuitively, the backdoored forward process integrates a trigger-related distribution shift $\rho_t\delta$ that gradually shifts the target image towards the trigger distribution $\delta$, with $\rho_t$ is the scale of the shift. This means that in the reverse process, the UNet must be trained to reverse the forward process by enforcing another trigger-related distribution shift $\widetilde{\rho}_t\delta$ that is in the reversed direction. Thus, backdoor attackers must preserve this distribution shift in every step along the Markov chain.

Based on this insight, An et al.\cite{an2024elijah} proposed a trigger inversion method that can find the trigger $\delta$ by analyzing the distribution shift of the UNet's output. The authors model the distribution shift as a linear dependence: $x_t^* - x_t = \lambda_t\delta$ and $x_{t-1}^* - x_{t-1} = \lambda_{t-1}\delta$, where $\lambda_t$ is a coefficient to model the trigger distribution shift. This means that if the UNet's input contains the trigger, the UNet's output must preserve the trigger distribution shift:
\begin{equation}
    \ex_{x_t}[M(x_t+\lambda_t\delta, t)] - \ex_{x_t}[M(x_t,t)] = \lambda_{t-1}\delta,
\end{equation}
where $M(\cdot)$ is the UNet. As a result, the problem of finding the backdoor trigger $\delta$ can be formalized as the following objective:
\begin{equation}
    \min_\delta \ex_t \left[ \| \ex_{x_t}[M(x_t+\lambda_t\delta, t)] - \ex_{x_t}[M(x_t,t)] - \lambda_{t-1}\delta \| \right].
\end{equation}

However, minimizing the above objective over all timesteps $t$ from $T$ to 1 is computationally inefficient. Therefore, the authors decided to only consider the step $T$, with $x_T \sim \mathcal{N}(0,\mathbf{I})$ and $x^*_T \sim \mathcal{N}(\delta,\mathbf{I})$. As a result, the objective is simplified to:
\begin{equation}
    \min_\delta \| \ex_{\epsilon\sim \mathcal{N}(0,1)}[M(\epsilon+\delta,T)] - \lambda\delta \|.
\end{equation}

Once a candidate trigger $\delta$ is figured out, we can verify whether the model is truly backdoored via the following observation: If a model is backdoored by a trigger $\delta$, sampling from $x_T^* \sim \mathcal{N}(\delta, \mathbf{I})$ consistently results in the target image. Otherwise, if the input noise is benign (i.e., $x_T \sim \mathcal{N}(0, \mathbf{I})$), the generation results would be very diverse due to stochastic factors of DMs during the reverse process. Therefore, by sampling $N$ times using the candidate trigger, then estimating a similarity score (e.g., KL Divergence, cosine similarity, total variance loss, or absolute distance) between different trials, one can confidently say that a DM is backdoored if the computed similarity is abnormally high. However, the main problem here is to choose a suitable ``similarity threshold" to determine if a DM is backdoored. 

The authors in\cite{an2024elijah} use total variance loss and absolute distance between different trials to compute the score for each DM, then construct a random forest consisting of hundreds of DMs (both clean and backdoored) to figure out a suitable score threshold. On the other hand, Guan et al.\cite{guan2024ufid} compute the cosine similarity between each pair of images generated from a sampled input, then constructing a similarity graph to assess the overall score for the input. If the overall score is higher than a predefined threshold, the input noise is considered a backdoor trigger. In this work, the threshold is calculated based on benign/clean samples. However, this work assumes that the trigger has been recognized earlier, and their method only verifies whether this candidate trigger is truly a backdoor trigger. 
This is not a reasonable assumption in practice, as inversing the trigger of a backdoored DM is often much harder than verifying a potential trigger.

In another work, Sui et al.\cite{sui2024disdet} argue that the distribution of a backdoor trigger should be significantly different from the distribution of a standard Gaussian noise. Therefore, the authors compute the KL Divergence between the UNet's input $x_T$ and a Gaussian noise, then choose a statistical threshold which is equal to the mean score plus three times of the score variance. However, using $x_T$ to compute the score might not be an efficient approach, as there are certain methods that can make the trigger almost invisible and resembles a normal noise (e.g., the reviewed framework in \cite{li2023learnable}). Furthermore, the work\cite{sui2024disdet} cannot inverse the trigger; it is just limited to verifying whether an input noise is a backdoor trigger.

Beyond backdoor detection, An et al.\cite{an2024elijah} also proposed a method to remove the backdoor effect embedded in a particular DM. This method purifies a backdoored DM so it will generate benign outputs even if the input is the backdoor trigger. To do so, the authors proposed a loss function that shifts the backdoored output's distribution towards the benign distribution. This loss function is the absolute distance between (i) the DM's output in presence of backdoored trigger and (ii) the output of the frozen DM when using benign input. Especially, this can be done without accessing the real data.

% \usepackage{multirow}
\begin{table*}[]
\centering
\caption{A summary of countermeasures for adversarial attacks on DMs. CA refers to cross-attention layers, while TE is text encoder.}
\label{table:adversarial-countermeasure}
\renewcommand{\arraystretch}{1.3} 
\begin{tabular}{|l|ccc|c|c|c|}
\hline
\rowcolor[HTML]{EFEFEF} 
\multicolumn{1}{|c|}{\cellcolor[HTML]{EFEFEF}} & \multicolumn{3}{c|}{\cellcolor[HTML]{EFEFEF}\textbf{Targeted Component}} & \cellcolor[HTML]{EFEFEF} & \cellcolor[HTML]{EFEFEF} & \cellcolor[HTML]{EFEFEF} \\ \cline{2-4}
\rowcolor[HTML]{EFEFEF} 
\multicolumn{1}{|c|}{\multirow{-2}{*}{\cellcolor[HTML]{EFEFEF}\textbf{Reference}}} & \multicolumn{1}{c|}{\cellcolor[HTML]{EFEFEF}\textbf{UNet}} & \multicolumn{1}{c|}{\cellcolor[HTML]{EFEFEF}\textbf{CA}} & \textbf{TE} & \multirow{-2}{*}{\cellcolor[HTML]{EFEFEF}\textbf{Guidance}} & \multirow{-2}{*}{\cellcolor[HTML]{EFEFEF}\textbf{Method}} & \multirow{-2}{*}{\cellcolor[HTML]{EFEFEF}\textbf{Strategy}} \\ \hline
ESD\cite{gandikota2023erasing} & \multicolumn{1}{c|}{\checkmark} & \multicolumn{1}{c|}{\checkmark} &  & CFG & Fine-Tuning & Unlearning \\ \hline
AC\cite{kumari2023ablating} & \multicolumn{1}{c|}{\checkmark} & \multicolumn{1}{c|}{} &  & Anchor Concept, Poisoning Dataset & Fine-Tuning & Unlearning \\ \hline
FMN\cite{zhang2024forget} & \multicolumn{1}{c|}{} & \multicolumn{1}{c|}{\checkmark} &  & Attention Maps & Fine-Tuning & Unlearning \\ \hline
ABO\cite{hong2024all} & \multicolumn{1}{c|}{\checkmark} & \multicolumn{1}{c|}{\checkmark} &  & Replacing Concept, CFG & Fine-Tuning & Unlearning \\ \hline
SLD\cite{schramowski2023safe} & \multicolumn{1}{c|}{} & \multicolumn{1}{c|}{} &  & CFG & Inference & Unlearning \\ \hline
EraseDiff\cite{wu2024erasediff} & \multicolumn{1}{c|}{\checkmark} & \multicolumn{1}{c|}{\checkmark} &  & Crafted Dataset & Retraining & Unlearning \\ \hline
Erasing\cite{fuchi2024erasing} & \multicolumn{1}{c|}{} & \multicolumn{1}{c|}{} & \checkmark & Reversed SGD & Fine-Tuning & Unlearning \\ \hline
DT\cite{ni2023degeneration} & \multicolumn{1}{c|}{\checkmark} & \multicolumn{1}{c|}{\checkmark} &  & Scrambled-Grid Dataset & Fine-Tuning & Unlearning \\ \hline
GuardT2I\cite{yang2024guardt2i} & \multicolumn{1}{c|}{} & \multicolumn{1}{c|}{} & \checkmark & Verbalizer, BERT & Prompt Verification & Red-Teaming \\ \hline
Pruning\cite{yang2024pruning} & \multicolumn{1}{c|}{\checkmark} & \multicolumn{1}{c|}{\checkmark} &  & Objective of ESD or AC & Model Pruning & Unlearning \\ \hline
MCU\cite{li2024machine} & \multicolumn{1}{c|}{\checkmark} & \multicolumn{1}{c|}{\checkmark} &  & KL Divergence & Fine-Tuning & Unlearning \\ \hline
SA\cite{heng2024selective} & \multicolumn{1}{c|}{\checkmark} & \multicolumn{1}{c|}{\checkmark} &  & EWC, GR & Fine-Tuning & Unlearning \\ \hline
RTSF\cite{rando2022red} & \multicolumn{1}{c|}{} & \multicolumn{1}{c|}{} & \checkmark & Hand-Crafted Prompt & Prompt Verification & Red-Teaming \\ \hline
SPM\cite{lyu2024one} & \multicolumn{1}{c|}{\checkmark} & \multicolumn{1}{c|}{\checkmark} &  & Anchor Concept, PEFT & Fine-Tuning & Unlearning \\ \hline
Receler\cite{huang2023receler} & \multicolumn{1}{c|}{\checkmark} & \multicolumn{1}{c|}{\checkmark} &  & CFG, PEFT, adversarial loss & Fine-Tuning & Unlearning \\ \hline
UC\cite{wu2024unlearning} & \multicolumn{1}{c|}{\checkmark} & \multicolumn{1}{c|}{\checkmark} &  & Anchor Concept, GAN discriminator & Fine-Tuning & Unlearning \\ \hline
RACE\cite{kim2024race} & \multicolumn{1}{c|}{\checkmark} & \multicolumn{1}{c|}{\checkmark} &  & Adversarial loss & Fine-Tuning & Unlearning \\ \hline
P4D\cite{chin2023prompting4debugging} & \multicolumn{1}{c|}{} & \multicolumn{1}{c|}{} & \checkmark & Adversarial noise & Prompt Generation & Red-Teaming \\ \hline
\end{tabular}
\end{table*}

\subsection{Countermeasures of Adversarial Attacks}
In general, the main purpose of adversarial attacks on DMs is to generate inappropriate content that is low-quality or unaligned with the provided text prompt. In certain circumstance, the generated content can be harmful, including NSFW concepts like racism, horror, politics, and violence. In general, there are three efficient approaches to eliminate such NSFW concepts from generated adversarial images, which are safety filter algorithms, red-teaming techniques, and MU\cite{bourtoule2021machine, cao2015towards}, which is also referred as concept erasing/forgetting in DMs. 

\subsubsection{Safety Filter and Red-Teaming Tools}
Safety filter is often deployed in various public DMs. It compares the latent representation of the generated image with pre-computed representations of multiple NSFW concepts. If the similarity score between the image's representation and any NSFW embedding is higher than a threshold, the generated image is considered NSFW and it is filtered out. However, it has been shown that these safety filters are still vulnerable and they can be bypassed easily by different methods, including both hand-crafted and learnable ones\cite{rando2022red, chin2023prompting4debugging}. On the other hand, red-teaming methods\cite{yang2024guardt2i, rando2022red, chin2023prompting4debugging} resist against adversarial attacks by proactively figuring out potential adversarial prompts that can make DMs generate NSFW content. In terms of red-teaming for DMs, Rando et al.\cite{rando2022red} reverse-engineer the stable diffusion's safety filter to identify its vulnerabilities in generating NSFW concepts. The authors find out that only sexual content is prevented by safety filters, while other NSFW concepts like violence, gore, and vulgar content are ignored. Furthermore, it has been shown that diluting input prompts by adding extra (irrelevant) details into the adversarial prompts can help bypassing safety filters as this disturbs the prompt's embedding, making the similarity score lower than the safety filter's threshold. These insights are important in figuring out potential adversarial prompts, thereby improving safety filters and designing stronger adversarial defense systems. Nevertheless, \cite{rando2022red} depend entirely on human handcrafting, which is not scalable. To enhance scalability, another red-teaming tool named P4D is proposed in\cite{chin2023prompting4debugging}, which is a learnable method that can find adversarial prompts by using a prompt-engineering approach. P4D finds an alternative prompt that is different from the target sensitive prompt, but the generated images are still similar to each other. This is done in the latent space, in which the alternative prompt is optimized by minimizing the different between (i) the noises predicted by the DM when using the original sensitive prompt as input, and (ii) the noise predicted when using the adversarial prompt. 

While the goal of \cite{rando2022red} and \cite{chin2023prompting4debugging} is to search for adversarial prompts, GuardT2I\cite{yang2024guardt2i} aims to verify whether a provided prompt is adversarial or not. In particular, GuardT2I analyzes the text embeddings of stable diffusion for prompt verification. First, it trains an autoregressive LLM that receives the DM’s text embeddings as conditions, then generates a prompt interpretation that convey the meaning of the text embedding. Then, the authors proposed two modules to check if the prompt is adversarial: (i) the first module is a verbalizer that checks if the interpreted prompt contains sensitive words; and (ii) the second module is a BERT-based sentence similarity checker that verifies whether the original prompt and the interpreted prompt is similar to each other. If the verbalizer finds out sensitive words, or the similarity checker shows that the two prompts are dissimilar, the prompt is considered NSFW. 

\subsubsection{Machine Unlearning for Concept Erasing}
Another efficient approach to tackle DM-targeted adversarial attack is MU-based concept erasing. Specifically, these techniques allow DMs to forget the knowledge of a particular concept such as nudity and violence, making it unable to generate these types of images no matter the provided prompt. This can be implemented via retraining, fine-tuning, or intervening in the inference. In terms of model retraining, Erasediff\cite{wu2024erasediff} divides the dataset into two parts: The first part contains the forgetting data, while the second part contains the remaining (normal) data.  It retrains the DM on the normal data with the same loss as standard DDPMs, which minimizes the ELBO. On the other hand, it fine-tunes the DM on the forgetting data with an opposite loss that maximizes the ELBO. This approach is limited in terms of scalability as it requires retraining the DMs on a large dataset.

ESD\cite{gandikota2023erasing} is among the first studies that investigate fine-tuning for concept erasing. The authors invert the sign of classifier-free guidance (CFG) to obtain a negative guidance term. This negative guidance is used to fine-tune DMs, moving the data distribution away from a specific condition $c$. As a result, it minimizes the probability that the generated images are labeled as $c$. The authors also proposed two fine-tuning strategies. The first strategy only fine-tunes cross-attention modules without intervening on other parameters. The second strategy only fine-tunes unconditional layers and keep cross-attention parameters unchanged. Experimental results showed that the first strategy only erases a specific content/style (e.g., Van Gogh) when its name is mentioned in the text prompt. On the other hand, the second strategy offers a more general and universal effect. For example, it removes “nudity” globally even in case this word is not mentioned in the prompt. Another framework called RACE\cite{kim2024race} is built on top of the training objective ESD\cite{gandikota2023erasing} with certain enhancement to improve the performance of concept erasing. In this method, the authors first simulate an adversarial attack that learns a perturbation $\sigma$ to make the targeted DM generates an unwanted erasing concept. This perturbation is integrated into the objective function of ESD to fine-tune the target DM. In the experiments, it has been shown that RACE achieves a better performance compared to ESD.

Also based on the principles of CFG, the authors in \cite{schramowski2023safe} proposed SLD, a framework that deploys concept erasing at the inference instead of fine-tuning. In addition to the normal CFG term, the authors add a safety guidance term that guide the DM not to generate an erasing concept that we aim to remove. This safety guidance term is similar to the CFG term, but the condition is such the erasing concept. This term is controlled by a scaling factor that provides a trade-off between removing inappropriate content and keeping the changes minimal for a high-quality generation result.

Receler\cite{huang2023receler} also use an objective that is based on negative guidance, which is similar to ESD. However, the authors introduce the following improvements: (i) the work applies parameter efficient fine-tuning (PEFT) to lower the cost of fine-tuning; (ii) it introduces a concept-localized regularization that helps ensuring locality of generated images on non-forgetting concepts; (iii) the erasing loss is trained concurrently against an adversarial loss that aims to generate adversarial examples. As a results, Receler is more robust than SLD and ESD in various experiments like nudity-erasing and adversarial attacks.
While ESD, SLD, and Receler use CFG in different ways to mislead DMs, in another work named FMN, Zhang et al.\cite{zhang2024forget} conduct concept erasing by applying a loss function that minimizes attention maps of the fine-tuned DMs.  However, FMN faces difficulties in forgetting abstract concepts. 

Instead of misleading DMs away from the erasing concept, the work called AC presented in\cite{kumari2023ablating} redirects the DM's distribution into another anchor concept that is more general than the original one. For example, the anchor concept of ``A Grumpy cat" is ``A cat". By making the targeted DM understand ``A Grumpy cat" as ``A cat", it can be said that the DM has forgotten the concept of ``Grumpy cat". To do so, the authors proposed two methods to fine-tune DMs. In the first method, a frozen DM is used with the anchor prompt ``A cat" to teach the fine-tuned DM, given that the input of the fine-tune DM is the original prompt ``A Grumpy cat”. The second method fine-tunes the DM with modified image-text data. For example, the text label is ``A Grumpy cat”, but the image is just a random cat (not Grumpy). Both methods make the fine-tuned DM treat ``Grumpy cat” as normal cats.

The combination of CFG and anchor concept is introduced by Hong et al.\cite{hong2024all} in a work named ABO. Similar to AC\cite{kumari2023ablating}, ABO also uses an anchor concept, with a frozen teacher DM to teach and fine-tune a student DM based on this concept. However, the anchor concept in ABO is defined by users instead of using a more general concept. This forms a concept loss term that makes the student DM forgets knowledge about the erasing concept. In addition, this paper also uses CFG to introduce a penalty loss term that provides unconditional feedback. As a result, this method outperforms previous work like ESD, AC, and FMN in various metrics like FID, CLIP Score, and Structural Similarity Index metric (SSIM).

The use of an anchor concept for concept erasing is also exploited in UC\cite{wu2024unlearning} and SPM\cite{lyu2024one}. SPM\cite{lyu2024one} employs parameter efficient fine-tuning (PEFT). The authors design a trainable lightweight adapter that can be inserted into different layers of the DM’s neural network. This adapter only has one dimension; thus, it can enable precise concept erasing with very low cost. To train this scheme, the authors proposed a latent anchoring method, in which a surrogate concept is chosen and combined with the target concept to fine-tune the model. The objective function thus consisting of two terms, an erasing loss and an anchoring loss, ensuring that the concept is erased while the impact on other retaining concepts is minimal. As a result, this framework achieves a better performance than previous studies like ESD, ConAbl, and SA in terms of FID score, while only lower than ESD in CS and CER. 
In UC\cite{wu2024unlearning}, the authors design a method for unlearning concepts that is inspired by GAN. Similar to SPM\cite{lyu2024one}, AC\cite{kumari2023ablating}, and ABO\cite{hong2024all}, UC also misleads the erasing concept to an anchor concept. This is done by using a GAN-based scheme in the latent space, in which a discriminator tries to distinguish if the predicted noise is from the erasing concept or from the anchor concept, while the DM is fine-tuned to ensure that the discriminator cannot distinguish this. It is formulated as a min-max objective that optimizes both the DM and the discriminator at the same time. The performance of this work is comparable to ESD in terms of concept unlearning, while it performs similarly to SPM regarding style unlearning. 

However, due to using an anchor concept, the presented methods necessitate a significant resource for preparing fine-tuning data. To facilitate data preparation, the authors in\cite{heng2024selective} proposed SA, which enables concept erasing by using conditional samples generated by such the DMs prior to fine-tuning. The loss function of SA is derived based on two continual learning techniques, Elastic Weight Consolidation (EWC) and Generative Replay (GR), which enables controllable forgetting and works across different types of generative models and conditioning schemes. As a result, it outperforms SLD and ESD in terms of classification-based scores like GCD. 

In terms of erasing robustness, the authors in\cite{yang2024pruning} argued that previous studies like ESD or AC just “deactivate” some neurons to remove the erasing concept. However, they can be reactivated if the input prompts are cleverly crafted. Therefore, this paper aims to prune/remove such neurons that associated with the erasing concept. First, the authors use an existing concept-erasing objective like ESD or AC to identity these neurons and estimate their sensitivity. With a predefined sensitivity threshold, a hard mask is derived so that it decides which neurons to prune. As a result, this method achieves a significantly higher performance than previous methods like ESD, FMN, and AC given that the adversarial strategy is UnlearnDiff\cite{zhang2023generate} and P4D\cite{chin2023prompting4debugging}.

However, most presented methods are only applicable for DMs. The authors in\cite{li2024machine} generalize the training objective to make it applicable for a wide range of generative models instead of only DMs. Specifically, the optimization goal is to minimize the KL divergence between the generated results of the retaining dataset (to ensure generation quality), while maximizing the KL divergence between those of the erasing dataset (to make the models forget that concept). However, since KL divergence between high-dimensional images is intractable, the authors proposed a more efficient method that convert KL divergence into a L2 loss, based on mutual information (MI). Since this objective is not specific to any model type, it can be used for various categories of generative models like VQ-VAE, MAE, and DMs.

Fine-tuning the denoising model is a popular approach for concept erasing. However, it might take significant time and computational resource in practice. The authors in\cite{fuchi2024erasing} instead fine-tune only the text encoder of DMs to achieve concept erasing. It simply updates the text encoder with a reverse direction of gradient descent, using several images of the target concept. Although this method is very simple, it improves the fine-tuning time significantly, which takes only around 8 seconds. However, the erasing robustness is limited than previous studies.

In another work\cite{ni2023degeneration}, the authors figure out two observations: (i) Low-frequency features (e.g., general shape, structure, and undetailed content) is the primary components of distribution distance, and they are steered by conditional information; and (ii) the semantic information of generated images are affected significantly by variations in the initial noise.
Thus, to make DMs forget unwanted concepts, the authors proposed to use scrambled grid to disrupt visually low-frequency features of the erasing concepts, forming a degraded dataset. Then, this dataset is used to fine-tune DMs.
As a result, it achieves a higher performance than SLD and ESD in terms of CLIP, FID, and IS scores.

\subsection{Countermeasures of Membership Inference Attacks}
MIAs pose a significant threat to the privacy of DMs trained on sensitive data. In order to mitigate these risks, several defense strategies have been proposed, including regularization, differential privacy, and knowledge distillation. They are presented as follows.

\subsubsection{Regularization} MIA methods primarily benefit from the overfitting tendency in deep learning models by leveraging the differences in their behaviors when processing member and non-member samples. To mitigate privacy risks associated with these attacks, a straightforward defense approach is to apply regularization techniques to enhance the generalization capability of the target model, thereby reducing the overfitting phenomenon and mitigating the model's vulnerability.
Specifically, Zhai et al.\cite{zhai2024membership} show that removing the default data augmentation techniques from the target model's training scripts makes it more vulnerable to their attack methods. In addition, both Duan et al.\cite{duan2023diffusion} and Pang et al.\cite{pang2023white} find that applying Cutout\cite{devries2017improved} and RandomHorizontalFlip techniques to the target model's training process causes their attack performance decrease to a certain degree.
However, when employing RandAugment\cite{cubuk2020randaugment} and $l_2$-regularization, \cite{duan2023diffusion} and \cite{pang2023white} observe that DMs fail to converge during training, resulting in generating extremely low-quality images.

\subsubsection{Differential Privacy} Applying differential privacy (DP)\cite{dwork2006calibrating} perturbation on training, the work in\cite{dwork2008differential} stands as a standard mechanism for defending against privacy attacks. In particular, DMs can be trained with differentially private stochastic gradient descent (DP-SGD)\cite{abadi2016deep}, where the gradients are clipped and noised  during training to diminish the models' memorization of individual samples, thereby avoiding leaking sensitive information from the training datasets.
However, in practice, applying DP-SGD often results in an unacceptable trade-off between data privacy and the model's utility. Specifically, \cite{pang2023black, tang2023membership, hu2023membership, carlini2023extracting}, and \cite{pang2023white} observe that employing DP-SGD during training causes DMs to fail to converge, which leads to a significant drop in the models' performance and the generation of meaningless images. Furthermore, incorporating DP into training requires much higher computational costs and time, posing a significant challenge when applying to large-scale DMs.

\subsubsection{Knowledge Distillation} The knowledge distillation strategy aims to transfer knowledge from a teacher model to a student model, enabling the student to attain performance comparable to that of the teacher. To this end, the outputs of the teacher model are used to train the student model, guiding it to mimic the behaviors exhibited by its teacher.
Exploiting this mechanism to reduce data privacy risks, Fernandez et al.\cite{fernandez2023privacy} propose Privacy Distillation, a framework that prevents text-conditional DMs from exposing identifiable information about their training data. To this end, they first train a private DM $\epsilon^{pri}_\theta$ on the real data $\textit{D}_{real}$. Then, captions from $\textit{D}_{real}$ are fed into $\epsilon^{pri}_\theta$ to generate images, forming a synthetic dataset $\textit{D}_{synth}$. Following this, the authors train a Siamese neural network to calculate the \textit{re-identification score} between generated images and real images, thereby filtering $\textit{D}_{synth}$ by removing the synthetic images that are considered identifiable. Finally, the filtered synthetic dataset is used to train a public DM $\epsilon^{pub}_\theta$ that can be shared openly and safely. Although Privacy Distillation significantly reduces re-identification risk by preventing public models from accessing real data and identifiable information, retraining large-scale DMs is computationally expensive.

There are also several approaches that can be applied to mitigate the data privacy concerns arising from MIAs. However, there remains a significant gap in dedicated research on defense methods specifically tailored to DMs.

% \section{DMs as a Mean for Attacks and Defenses}\label{section:DM-attack-defense}

% \subsection{Diffusion-Based Attacks}

% \subsection{Diffusion-Based Defenses}

\section{Open Challenges and Research Directions}\label{section:open-challenges}
As discussed earlier, various studies regarding attacks and defenses for DMs have been proposed and cover three main branches of DM security. However, it can be seen that this field is still in its infancy, since many other issues of both attacks and countermeasures for DMs have not been investigated. Thus, this section presents a throughout discussion about open challenges and future research directions of DM security, inspiring future development towards more secure and privacy-preserving DM-based systems.

\subsection{Backdoor Attack}
In backdoor attacks, the attackers must modify the diffusion processes, loss function, and training data to embed the triggers into the target DMs. As a result, the backdoor can be activated when the backdoor trigger is fed into the model's input. However, the modifications applied on the backdoored DM eventually led to a significant downgrade on the model's performance when non-trigger inputs are used. This not only impacts user experience, but also decreases the stealthiness of the backdoor attack if the generative results of the backdoored DM are compared to those generated by a benign model.

Another limitation of current backdoor methods is that they are only applicable for image-based applications. It is known that DMs are also applied in other important domains like natural language processing (NLP)\cite{austin2021structured, hoogeboom2021argmax, li2022diffusion, savinov2021step, yu2022latent, lin2023text}, audio processing\cite{chen2020wavegrad, popov2021grad}, 3D generation\cite{xu2023dream3d, truong2024text, poole2022dreamfusion, lin2023magic3d, haque2023instruct, luo2021diffusion}, bioinformatics\cite{xu2022geodiff, luo2022antigen}, and time series tasks\cite{tashiro2021csdi, yan2021scoregrad, rasul2020multivariate}. Therefore, investigating backdoor attacks for these types of DMs is also a potential research direction. For example, one might try to backdoor an audio-based DM so that the target DM would generate harmful auditory content when the backdoor trigger is activated.

In terms of future research direction, embedding multiple backdoor triggers into a single DM can be a potential method to make backdoor attacks resist to backdoor defense methods like trigger inversion and trigger purification. Even if trigger inversion is applied to find out a backdoor trigger then purify it, the attackers can still use other embedded triggers to activate the backdoor target. In more general cases, we can also design various variations of backdoor attacks such as multi-trigger single-target, single-trigger multi-target, and multi-trigger multi-target, and multi-modal backdoor attacks.

\subsection{Backdoor Defense}
Trigger inversion is the foremost and also the most challenging stage of countering backdoor attacks. Once figuring out some candidate backdoor triggers via trigger inversion, it is more simple to verify whether these candidates are truly backdoor triggers based on various reviewed methods\cite{an2024elijah, guan2024ufid, sui2024disdet}. However, existing work only investigates backdoor inversion for image-based trigger\cite{an2024elijah}, while neglecting another important case in which the trigger is embedded into the text prompt. Therefore, investigating robust techniques to inverse textual triggers is an important future research direction. However, text-based trigger inversion is often more challenging since there are various strategies to embed textual triggers into the text prompt, including word/character appending, prepending, and replacing. In addition, unlike image-based triggers, textual triggers are discrete and could not be learnt by traditional gradient descent methods. 

Furthermore, current backdoor countermeasures only consider single-trigger single-target attacks. In future work, it is possible that multiple triggers can be embedded into a DM at the same time, making trigger inversion more challenging. Therefore, investigating defense methods that can detect thoroughly all the triggers is an important research direction.

\subsection{Adversarial Attack}
In existing studies, adversarial attack is usually carried out by adding a small perturbation into one of three target modalities, including input image\cite{salman2023raising, yu2024step, shan2023glaze, zhang2023robustness}, text prompt\cite{yang2023mma, yang2024sneakyprompt, zhuang2023pilot, gao2023evaluating, zhang2024revealing, liu2023discovering, liu2023riatig, kou2023character, zhang2023generate}, and fine-tuning images\cite{liang2023adversarial, liang2023mist, zhu2024watermark, van2023anti}. It has been shown that conducting adversarial attacks on different input modalities brings different extents of attacking performance, learning time, computational resource, and image quality. As the adversarial goals (e.g., image degradation or content shifting) are similar between different target modalities, perturbing multiple input modalities at the same time (e.g., both input image and text prompt) can potentially offer a higher performance. However, none of existing studies investigate attacking DMs concurrently from multiple modalities. This can be investigated in future work, where, for example, the perturbations are added to both the text prompt and input image to achieve a higher attacking performance.

In addition, current studies mostly focus on image-generation adversarial attacks with textual guidance, while neglecting other modalities/ data types such as audio, time series, and 3D models. Attacking such modalities requires more sophisticated strategies to learn the adversarial perturbation, since it involves additional techniques for these specific applications (e.g., DM-based 3D generation often requires a specific neural radiance fields model for 3D synthesis).

\subsection{Adversarial Defense}
In terms of countermeasures against adversarial attacks, most existing studies investigate concept erasing as a mean of defense. This method can solve NSFW generation by making the fine-tuned DMs forget about NSFW concepts, thus protecting users from harmful content even when the adversarial perturbation is fed into the DMs' input. However, countermeasures for other adversarial threats such as quality diminishing, concept shifting, and style transferring have not been investigating in existing work. In such cases, the purpose of attack is more general and not focuses on any particular concept, making it especially challenging to remove the adversarial effect. Hence, this remains an open challenge and necessitate further effort on future work. 
% most recent studies focus on concept erasing. It can solve only NSFW content generation, while other adversarial purposes like content shifting, quality diminishing can not be tackled.

\subsection{Membership Inference Attack}
Nowadays, pre-trained large-scale DMs are publicly available and easily accessible on the Internet. Fine-tuning these models for downstream tasks has become a very popular paradigm since it requires less computational resources compared to training a model from scratch. There are many new fine-tuning techniques adapted to DMs that are widely used, such as LoRA \cite{hu2021lora} and DreamBooth \cite{ruiz2023dreambooth}, which show the effectiveness of guiding the DM's generation process. Many show that fine-tuning large-scale DMs on small datasets makes them easily overfit, hence increasing the vulnerability to MIAs. Having said that, no studies have yet investigated the data privacy implications of fine-tuned DMs. This remains an open challenge for the research community.

In addition, while the shadow training technique \cite{shokri2017membership} is commonly used effectively in MIAs, its application to large-scale DMs demands intensive computational resources. This requirement poses a significant challenge to training shadow models for efficient MIA methods.

\subsection{Membership Inference Defense}
There are various adaptation techniques for DMs, including LoRA\cite{hu2021lora}, DreamBooth\cite{ruiz2023dreambooth}, and Textual Inversion \cite{gal2022image}. These techniques are often trained on small datasets as a supplementary module to guide the main models' generation styles. Due to their effectiveness and accessibility, they are used and shared widely on the Internet, potentially including misused modules trained on private or non-copyrighted data, such as a module developed to mimic the art style of a famous artist. Unfortunately, their ease of overfitting makes them more vulnerable to MIAs. Consequently, the data privacy risks of applying these adaptations require proper countermeasure methods.

Moreover, knowledge distillation is a well-established strategy widely used to mitigate data privacy risks for ML models against MIAs. Although numerous studies have successfully applied knowledge distillation to DMs, this approach has yet to be explored as a defense mechanism against such MIAs.

% \subsection{Attacks on DM-Based Security Systems}
% While DMs are well-known in image generation tasks, they can also be applied for various security applications such as adversarial purification\cite{nie2022diffusion, wang2022guided, wu2022guided, wu2023defending, silva2023diffdefense, sun2022pointdp, sun2023critical} and robustness certification\cite{carlinicertified}. However, these DM-based security systems can be manipulated by attacking such integrated DMs. This type of attack is often specific based on the properties of the targeted security systems. While several studies\cite{li2023change, kang2024diffattack} have investigated attacking strategies for security-centric DMs, countermeasures for these attacks are still neglected, making this an open and worth researching topic.

\section{Conclusion}\label{section:conclusion}
In this paper, we have provided a comprehensive survey of DMs regarding the security aspect. The paper begins with crucial background knowledge of different types of DMs including DDPMs, DDIMs, NCSNs, SDE, and multi-modal conditional models. In terms of attacks on DMs, they are categorized into three main groups, including backdoor attack, adversarial attack, and MIA. Each type of attack is surveyed comprehensively based on state-of-the-art attack methods presented in recent studies. Then, we emphasized the importance of DM's safety by providing an extensive survey on countermeasures of these attacks. Finally, we outlined current open challenges in DM's security and envisioned potential avenues for future research in this topic. Based on the review, we found out that most research works focus on only vision and language domains, while other modalities such as audio and time series were neglected in all three main types of attack. In addition, countermeasures for these DM-targeted attacks are still limited and needed further research to enhance the security aspect of DMs.

\footnotesize{
\bibliographystyle{IEEEtran}
\bibliography{reference.bib}
}

\end{document}